\documentclass{amsart}
\usepackage[margin=1in]{geometry}
\usepackage{amsmath, amsfonts, amssymb}
\usepackage{color}
\usepackage{algpseudocode}
\usepackage{algorithm}
\usepackage{siunitx}
\usepackage{graphicx}
\usepackage{subfig}
\usepackage{physics}
\usepackage{bbm}
\usepackage{tikz}
\usepackage{mathtools}
\usepackage{mathrsfs}

\usepackage{hyperref}
\usepackage{cleveref}

\newcommand{\dif}{\mathrm{d}} 
\newcommand{\dk}{\Delta k}
\newcommand{\dt}{\Delta t}
\newcommand{\e}{\mathrm{e}} 
\newcommand{\id}{\mathrm{id}} 
\newcommand{\ii}{\mathrm{i}} 
\newcommand{\sgn}{\mathrm{sgn}}
\DeclareMathOperator{\SPAN}{span}

\newtheorem{theorem}{Theorem}
\newtheorem{lemma}[theorem]{Lemma}

\newtheorem{proposition}[theorem]{Proposition}
\newtheorem{definition}[theorem]{Definition}

\theoremstyle{remark}
\newtheorem*{remark}{Remark}

\title{Differential Equation Based Path Integral for System-Bath Dynamics}

\author{Geshuo Wang}
\address[Geshuo Wang]{Department of Mathematics, National University of Singapore,
  Level 4, Block S17, 10 Lower Kent Ridge Road, Singapore 119076}
\email{geshuowang@u.nus.edu}

\author{Zhenning Cai}
\address[Zhenning Cai]{Department of Mathematics, National University of Singapore,
  Level 4, Block S17, 10 Lower Kent Ridge Road, Singapore 119076}
\email{matcz@nus.edu.sg}

\thanks{Zhenning Cai's work was supported by the Academic Research Fund of the Ministry of Education of Singapore under grant R-146-000-291-114.}

\keywords{Open quantum system, DEBPI, i-QuAPI}

\begin{document}

\maketitle

\begin{abstract}
We propose the differential equation based path integral (DEBPI) method to simulate the real-time evolution of open quantum systems. 
In this method, a system of partial differential equations is derived based on the continuation of a classical numerical method called iterative quasi-adiabatic propagator path integral (i-QuAPI). 
While the resulting system has infinite equations, we introduce a reasonable closure to obtain a series of finite systems. 
New numerical schemes can be derived by discretizing these differential equations. 
It is numerically verified that in certain cases, by selecting appropriate systems and applying suitable numerical schemes, the memory cost required in the i-QuAPI method can be significantly reduced.
\end{abstract}

\section{Introduction}
In physics, an open quantum system refers to a quantum system coupled with an external environment (bath). In reality, no quantum system can be absolutely isolated, and therefore 
the study of open quantum system is of great importance
 and nowadays has wide applications
 in the fields including quantum optics \cite{carmichael2009open}, 
 chemical physics \cite{weiss2012quantum,leggett1987dynamics},
 quantum information \cite{nielsen2002quantum}
 and even social sciences \cite{khrennikova2014application}.
In general, for an open quantum system,
 the large number of degrees of freedom in the bath
 prevents us from directly computing the density matrix
 of the coupled system. To carry out simulations, the bath is usually assumed to be some analytically solvable models such as harmonic oscillators so that 
 one can represent the quantum dynamics using path integrals, and
 the influence of the bath on the system of interest
 can be reduced to an influence functional \cite{feynman1963theory}.

Despite such simplification,
 the difficulty introduced by the system-bath coupling still exists
 because the influence functional couples the states of the system
 at any two time points,
 leading to a non-Markovian process.
Such process is often formulated using an integro-differential equation 
(known as Nakajima-Zwanzig equation), 
where the effect of the bath is expressed as a memory kernel, represented by the Nakajima-Zwanzig operator \cite{nakajima1958quantum, zwanzig1960ensemble}.
The numerical methods based on the computation of path integrals,
 such as quasi-adiabatic propagator path integral (QuAPI) \cite{makri1992improved} and hierarchical equations of motion (HEOM) \cite{tanimura1989two,strumpfer2012open} methods,
 require significant memory cost to
 store the contributions from different paths.
To reduce the storage requirement,  one idea is to make use of the decaying property of the influence functional to develop iterative schemes.
One typical method is the iterative QuAPI (i-QuAPI) method \cite{makri1995numerical,makri1998quantum},
 which truncates the influence functional to achieve finite memory length, so that partial summation can be carried out in the path integral.
Other approaches based on this idea  include the blip-summed method \cite{makri2014blip,makri2017blip},
 small matrix decomposition of the path integral (SMatPI) method \cite{makri2020small},
 and important path sampling \cite{sim1996tensor} design algorithms, which explore extra the numerical sparsity of the problem to further reduce the memory cost.
Some stochastic methods,
 such as diagrammatic quantum Monte Carlo (dQMC) method \cite{werner2009diagrammatic} and
 inchworm Monte Carlo method \cite{chen2017inchworm,cai2020inchworm,yang2021inclusion,cai2020numerical},
  apply stochastic approaches to estimate
 path integrals. These methods no longer suffer from the curse of memory cost.
However, numerical sign problem may appear for long-time simulations \cite{loh1990sign,cai2020numerical}.
There are also some modeling techniques to simplify the Nakajima-Zwanzig equation, 
 among which the Lindblad equation \cite{lindblad1976generators,gorini1976completely} considers the weak coupling limit, so that the memory effect can be ignored and the master equation becomes Markovian. 
Other approaches based on computable master equations include the post-Markovian master equation \cite{shabani2005completely} and the generalized quantum master equation \cite{Kelly2013efficient}.

Despite the fast progress in the numerical computation of open quantum systems, the i-QuAPI method remains as one of the fundamental and reliable methods, and is often used to compute reference solutions in the development of other methods. 
Since the strongest limitation of i-QuAPI is its memory cost, which grows exponentially with $1/\dt$, the present paper follows the basic idea of the i-QuAPI method and tries to break such exponential relationship by developing the differential equation based path integral (DEBPI) method. The DEBPI method formulates a system of differential equations such that i-QuAPI can be considered as its numerical scheme.
Afterwards, we can re-design the numerical scheme by discretizing the PDE system. It is numerically verified that in certain cases, the memory cost can be significantly reduced.

Throughout this paper, we will mainly focus on a simple open quantum system consisting of a flipping spin and a thermal bath.
Despite the simple setting of the system, it contains all the difficulties related to the system-bath coupling in general open quantum systems.
We refer the authors to \cite{vacchini2010exact} for some detailed discussion on this model.

For the rest of this paper,
 we present the derivation of DEBPI method and some numerical tests.
\Cref{Spin_Boson_And_IQUAPI} is a brief introduction
 to the spin-boson model and the i-QuAPI method.
In order to derive the continuous formulation
 for i-QuAPI method,
 we give a representation of the continuation of path segments and introduce notations for the relative quantities in \Cref{Sec_Paths}.
With all the preparation in \Cref{Spin_Boson_And_IQUAPI,Sec_Paths},
 we derive a governing partial differential equation system
 and discuss the required boundary condition in \Cref{sec:quapi_pde}. 
In addition, as the partial differential equation system includes infinite equations,
 we also provide the idea of dimensional truncation
 and the closure of the system in \Cref{sec:quapi_pde}.
In \Cref{Sec_Numerical_Experiments},
 we carry out some numerical experiments
 based on finite difference discretization method for DEBPI method
 and compare our results with i-QuAPI method.
In \Cref{Discussion_Conclusion},
 we outline the application of the idea in other finite state models
 and give a simple conclusion.
The derivation of i-QuAPI method is given in the appendix for reference.

\section{Spin-boson model and iterative QuAPI method}
\label{Spin_Boson_And_IQUAPI}
In a quantum system with Hamiltonian $H$,
the density matrix $\rho(t)$ 
satisfies the von Neumann equation
\begin{equation*}
	\ii\hbar \frac{\dif}{\dif t}\rho(t) = H\rho(t) - \rho(t)H,
	\label{vonNeumann}
\end{equation*}
where $\ii$ is the imaginary unit, and
 $\hbar$ is the reduced Planck's constant.
For simplicity,
 $\hbar$ is set to be 1
 in this paper.
The solution to the von Neumann equation
 can be formally written as $\rho(t) = \e^{-\ii H t} \rho(0) \e^{\ii H t}$.

For an open quantum system, the operators are defined on the tensor product space $\mathcal{H}_s \otimes \mathcal{H}_b$, where $\mathcal{H}_s$ and $\mathcal{H}_b$ represent the Hilbert spaces for the system and the bath, respectively.
 In general, the Hamiltonian $H$ has the form
\begin{equation}
    H = H_s \otimes \id_b + \id_s \otimes H_b + H_{sb}
    \label{HamiltonianSpit}
\end{equation}
where $\id_s, \id_b$ are the identity operators
 on $\mathcal{H}_s$
 and $\mathcal{H}_b$, respectively.
The operators $H_s$ and $H_b$ are the Hamiltonian operators for the system and the bath without coupling,
and the last term $H_{sb}$ describes the system-bath interaction,
 which lead to entanglement between system and bath.
For simplicity, below we focus on the study of a specific case of the open quantum system, where the system contains a single spin. The model and the iterative QuAPI method will be introduced in the following subsections.

\subsection{Spin-boson model}
A fundamental example of open quantum systems
 is the spin-boson model
 where the Hilbert spaces are
\begin{equation*}
    \mathcal{H}_s = \SPAN\{\ket{-1},
    \ket{+1}\},
    \quad
    \mathcal{H}_b = \bigotimes_j
    (L^2 (\mathbb{R}^3))
\end{equation*}
with $L^2(\mathbb{R}^3)$ being the $L^2$ space over $\mathbb{R}^3$
 and $j$ is the index of harmonic oscillators.
In spin-boson system, there are two energy levels in the energy set $\mathcal{S}=\{-1,+1\}$.
In such a system,
 the Hamiltonians in (\ref{HamiltonianSpit})
 can be represented by
\begin{equation}
    H_s = \epsilon \hat{\sigma}_z
    + \Delta \hat{\sigma}_x,
    \quad
    H_b = \sum_j \frac{1}{2}
    (\hat{p}_j^2 + \omega_j^2 \hat{q}_j^2),
    \quad
    H_{sb} = \hat{\sigma}_z \otimes \left(
    \sum_j c_j \hat{q}_j
    \right)
    \label{Hamiltonian}
\end{equation}
where $\hat{\sigma}_x, \hat{\sigma}_z$
 are Pauli matrices, and
 $\hat{p}_j$ and $\hat{q}_j$ are momentum operator
 and position operator of the $j$-th harmonic oscillator, respectively.
$\epsilon$ represents the energy difference
 between two spin states
and $\Delta$ is the frequency of the spin flipping.
The parameter $\omega_j$ is the frequency of the $j$-th harmonic oscillator
 while $c_j$ is the coupling intensity between the spin and the $j$-th harmonic oscillator.

For convenience, it is supposed that
 the system and bath are not coupled initially,
 which can be expressed mathematically by
\begin{equation*}
	\rho(0) = \rho_s(0) \otimes \rho_b(0)
\end{equation*}
with $\rho_s$ and $\rho_b$
 representing the density matrices
 of the system and the bath, respectively.
The reduced density matrix,
 which is the density matrix 
 of the system,
 can be represented by the partial trace 
 of the whole density matrix:
\begin{equation*}
	\rho_s (t) = \tr_b \rho(t),
\end{equation*}
 where $\tr_b$ is the partial trace operator with respect to $\mathcal{H}_b$.
In the spin-boson model, this reduced density matrix $\rho_s$ is essentially a $2\times 2$ matrix.
The algorithm discussed in this paper
 can be generalized to more complicated model.
However, in this paper,
 only spin-boson model is involved
 and there is a simple discussion about other finite state models
 in \Cref{Discussion_Conclusion}.

\subsection{Iterative QuAPI method}
To compute the reduced density matrix
 in a spin-boson model,
 the iterative quasi-adiabatic propagator path integral
 (i-QuAPI) method is developed in \cite{makri1995numerical,makri1998quantum}. It is assumed that the density matrix has a finite memory length $T$ although it is non-Markovian. Then we choose the time step $\Delta t$ such that $\Delta k := T/\Delta t$ is an integer. The density matrix can be obtained through computing the maps $A_l: \mathcal{S}^{2\dk} \rightarrow \mathbb{C}$ iteratively:
\begin{align}
	&A_0(S_{\dk-1}, \cdots, S_0) := \prod_{k_1=0}^{\dk-1} \prod_{k_2=0}^{k_1}
		I(S_{k_1}, S_{k_2}) 
		\mel{s_0^+}{\rho_s(0)}{s_0^-};
	 \label{initial}\\
	&\Lambda(S_k,\cdots, S_{k-\dk}) := \prod_{m=0}^{\dk} I(S_k, S_{k-m});
	\label{propagator}\\
	&A_{k-\dk+1} (S_{k},\cdots, S_{k-\dk+1})
	:= \sum_{S_{k-\dk}\in \mathcal{S}^2}
	\Lambda(S_k,\cdots, S_{k-\dk})
	A_{k-\dk}(S_{k-1},\cdots,S_{k-\dk})
	\label{propagation}\\
	&\qquad\text{for } k = \dk,\cdots, N-1.\notag
\end{align}
where $S_j = (s_j^+, s_j^-)$
 represents the states on both branches of the path
 and $I(S_j, S_{j'})$ is defined to be
\begin{equation}
	I(S_j, S_{j'}) = \begin{cases}
		\e^{- (s_j^+ - s_j^-)(\eta_{j,j'}s_{j'}^+ - \eta_{j,j'}^* s_{j'}^-)},\quad j-j'\not= 1 \\
		\mel{s_j^+}{\e^{-\ii H_0 \dt}}{s_{j-1}^+}
		\mel{s_{j-1}^-}{\e^{\ii H_0 \dt}}{s_j^-}
		\e^{- (s_j^+ - s_j^-)(\eta_{j,j'}s_{j'}^+ - \eta_{j,j'}^* s_{j'}^-)},\quad j-j' = 1 
	\end{cases}
	\label{definitionI}
\end{equation}
where $\eta^*$ is the complex conjugate of $\eta$.
The complex function $\eta$ has different forms for different $j$ and $j'$ \cite{makri1995tensor1}.
 For example,
 if only positive frequencies are considered,
 when $0<j'<j<N$, 
\begin{equation}
    \eta_{j,j'} = \frac{4}{\pi}\int_0^\infty\dif \omega
    \frac{J(\omega)}{\omega^2}
    \sin^2\left(\frac{\omega\dt}{2}\right)
    \left(\coth\left(\frac{\beta\omega}{2}\right)
    \cos(\omega\dt(k-k'))
    -\ii\sin(\omega\dt(k-k'))\right).
    \label{eta}
\end{equation} 
In the expression of $\eta$, 
\begin{equation*}
	J(\omega) 
	= \frac{\pi}{2}\sum_{j}
	\frac{c_j^2}{m_j\omega_j}\delta(\omega-\omega_j)
	\label{J}
\end{equation*}
and $\delta$ is the Dirac delta function.
The derivation of the i-QuAPI method
 can be found in the appendix.
With the values of $A_{k-\dk}(S_k,\cdots,S_{k-\dk})$,
 the entries in the reduced density matrix
 can be computed by
\begin{equation}
	\rho_s(N\dt) = \sum_{S_{N-\dk+1},\cdots,S_{N}\in\mathcal{S}}
	A_{N-\dk}(S_N,\cdots,S_{N-\dk+1})
	\ket{s_N^+} \bra{s_N^-}.
	\label{densityMatrixDiscrete}
\end{equation}

The system-bath coupling leads to a non-Markovian process
 where the evolution of the system density matrix depends on the complete history, and the intensity of the dependence is determined by the discrete bath response function $\eta_{j,j'}$.
The method of iterative quasi-adiabatic propagator path integral (i-QuAPI),  however, makes use of the decaying property
 of the discrete bath response function $\eta_{j,j'}$ and truncate the memory length to a fixed and finite time $T$, so that the evolution is similar to the time-delay systems, making the simulation more feasible.

The major obstacle of this method is the memory constraint.
In each step, 
 there are in total $2^{2\dk}$ different values of $A_l$ to be stored in the memory.
As $\dk$ is defined to be $T/\dt$,
 and $T$ is fixed according to the problem itself,
 the restriction on $\dk$ directly results in
 a lower bound of $\dt$.
The contradiction is that for more accurate results,
 smaller $\dt$ should be chosen,
 which is prohibited by the memory.
For example, if the truncation time $T$ is 1.5 and the time step $\Delta t$ is chosen to be $0.1$, then $\dk = 15$ and the memory cost is approximately $2^{2\times 15} \times (16 \text{ bytes}) = 16\text{GB}$, where ``16 bytes'' stands for the size of a double-precision complex number.
However, if a smaller time step, for example, $\dt = 0.075$, is required,
the memory cost will become $2^{2\times 20} \times (16 \text{ bytes}) = 16 \text{TB}$, which is beyond the capacity of most workstations.

\section{Continuation of the Path Segments}
\label{Sec_Paths}
In order to cut off the memory cost, the most straightforward idea is to reduce the number of path segments that need to be stored. 
Equation \eqref{definitionI} shows that when the flipping rate $\Delta$ is small, the path segments with many spin flips may have little contribution to the density matrix, and therefore may be dropped in the path integrals without much loss of numerical accuracy.
To achieve this, we will carry out the following:
\begin{enumerate}
\item Find the governing equations of i-QuAPI by taking the limit $\Delta t \rightarrow 0$;
\item Truncate the small terms and provide a reasonable closure;
\item Discretize the resulting system partial differential equations to get a scheme with less memory cost.
\end{enumerate}
In this section, we will focus only on the representation of the unknown function which stands for a path segment with memory time $T$. The complete form of the governing equations will be derived in \Cref{sec:quapi_pde}.

\subsection{Representation of the paths}
\label{RepresentationOfPaths}
In the i-QuAPI method, each $A_l$ has a set of
 parameters $S_{k-1},\cdots,S_{k-\dk}$,
 representing a path segment in the path integral.
While the sequence $S_{k-1},\cdots,S_{k-\dk}$ represents the discrete path segment, we can regard it as a piecewise constant path segment of the continuous time 
 $s_{\dt}:[0,T)\rightarrow \mathcal{S}^2$,
 defined by
\begin{equation}
    s_{\dt}(\tau) = S_{k-j}, \quad \forall \tau \in \Big[(\dk-j)\Delta t, ~~ (\dk-j+1)\Delta t \Big),
    \quad j = 1,\cdots,\dk,
    \label{DefSdt}
\end{equation}
where $\Delta t = T/\dk$. 
More generally, we would like to consider any path segment $h: [0,T) \rightarrow \mathcal{S}^2$ with finite discontinuities, where each discontinuity denotes a flip of the classical spin. 
We assume that there are $D$ spin flips in the path segment $h$. 
To define $h$,
 we introduce a $D$-dimensional open simplex $\triangle_t^{(D)}$:
\begin{equation}
    \triangle_t^{(D)} = \left\{(\tau_1,\cdots,\tau_D) \,\bigg\vert\, \tau_1 > 0, \cdots, \tau_D > 0; ~~ \sum_{k=1}^D \tau_k < t \right\}
    \label{DefnSimplex}
\end{equation}
so that $h$ has the form
\begin{equation} 
h(\tau) = \left\{ \begin{array}{ll}
  (r_0^+, r_0^-), & \text{if } \tau \in [0,\tau_1), \\
  (r_d^+, r_d^-), & \text{if } \tau \in \left[\sum_{k=1}^d \tau_k, \sum_{k=1}^{d+1} \tau_k \right), \quad d = 1,\cdots,D-1, \\
  (r_D^+, r_D^-), & \text{if } \tau \in \left[\sum_{k=1}^D \tau_k, T \right),
\end{array} \right.
\label{contPath}
\end{equation}
where $(\tau_1, \cdots, \tau_D) \in \triangle^{(D)}_T$. In $h$, we allow the locations of the jumps to be anywhere in $(0,T)$ instead of only being the multiples of $\Delta t$. For the paths $s_{\dt}$ and $h$, we use $s_{\dt}^+$ ($s_{\dt}^-$) and $h^+$ ($h^-$) to denote their positive (negative) branches.

With these definitions, it is natural to express the quantity $A_{k-\dk}(S_{k-1},\cdots,S_{k-\dk})$ as $A((k-\dk)\dt, s_{\dt})$, 
where the first argument $(k-\dk)\dt$ plays the role of the subscript $(k-\dk)$, denoting the starting time of the path segment.
This can then be further extended to continuous time by writing the quantity as $A(t,h)$ for any $t \geqslant 0$ and $h$ holding the form \eqref{contPath}.
However, such a quantity $A(t,h)$ includes a non-classical argument $h$, which is inconvenient for the formulation of differential equations
and the design of numerical scheme. We would therefore like to find a more explicit form of $A(t,h)$ containing only real parameters.
 With the assumption that $h$ contains only finite discontinuities, the path segment can be fully determined by the following set of parameters:
\begin{itemize}
	\item $(r^+,r^-)$: the state of the path at time $t$, \textit{i.e.} the value of $h(0) = (h^+(0),h^-(0)) = (r_0^+,r_0^-)$.
	\item $D$: the number of spin flips (discontinuous points) in $h$.
	\item $\sgn$: a list of $D$ signs
	representing the branch of each spin flip.
	Each sign is either ``$+$'' (representing the positive branch $h^+$) or ``$-$'' (representing the negative branch $h^-$). 
    In addition, we introduce an augmenting operation by adding a sign
	 at the beginning of $\sgn$.
	For example, if $\sgn = [+,-,+,+]$,
	 then $[+,\sgn] := [+,+,-,+,+]$.
	\item $(\tau_1,\cdots,\tau_D)\in \triangle^{(D)}$: a sequence of times with the following meaning:
	\begin{equation*}
		\tau_k = \begin{cases}
			\text{time difference of the first spin flip and the starting time $t$}, \text{ if } k = 1; \\
			\text{time difference of the $k$-th flip and the $(k-1)$-th flip}, 
			\text{ if } k = 2,\cdots,D.
		\end{cases}
	\end{equation*}
\end{itemize}


For example, suppose the truncation time is $T = 4$ and the graph of the path segment $h$ is given in  \Cref{fig:demonstrationOfh}.
 Then the corresponding parameters are
\begin{itemize}
    \item $(r^+,r^-) = (h^+(0), h^-(0)) =  (+1,-1)$;
    \item $D = 5$ (total number of spin flips, 2 in positive branch and 3 in negative branch);
    \item $\sgn = [+,-,-,+,-]$ (branches of the spin flips);
    \item $[\tau_1,\tau_2,\tau_3,\tau_4,\tau_5]
    = [1.0,0.5,0.5,0.5,1.0]$ (indicating the times of the spin flips are $1.0$, $1.5$, $2.0$, $2.5$, $3.5$ from $t$).
\end{itemize}
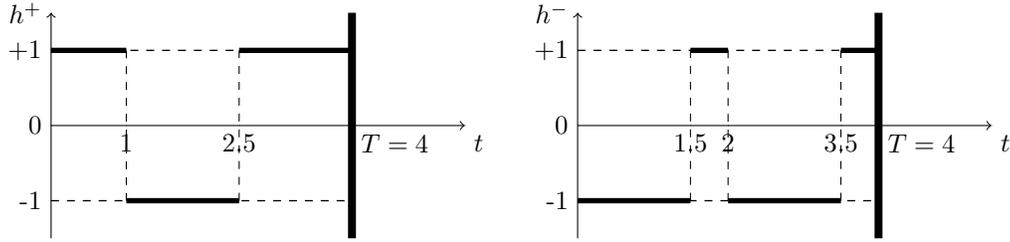
\begin{figure}[!ht]
    \centering
\begin{tikzpicture}
  \draw[->] (0,0) -- (5.5,0);
  \draw[->] (0,-1.5) -- (0,1.5);
  \draw[line width = 3pt] (4,-1.5) -- (4,1.5);
  \draw[line width = 2pt] (0,1) -- (1,1);
  \draw[dashed] (1,1) -- (1,-1);
  \draw[line width = 2pt] (1,-1) -- (2.5,-1);
  \draw[dashed] (2.5,-1) -- (2.5,1);
  \draw[line width = 2pt] (2.5,1) -- (4,1);
  \draw[dashed, line width=0.5pt] (0,1) -- (4,1);
  \draw[dashed, line width=0.5pt] (0,-1) -- (4,-1);
  
  \filldraw[black] (0,0) circle node[anchor=east] {0};
  \filldraw[black] (1,0) circle node[anchor=north] {1};
  \filldraw[black] (2.5,0) circle node[anchor=north] {2.5};
  \filldraw[black] (4,0) circle node[anchor=north west] {$T=4$};
  \filldraw[black] (5.5,0) circle node[anchor=north west] {$t$};
  \filldraw[black] (0,1) circle node[anchor=east] {+1};
  \filldraw[black] (0,-1) circle node[anchor=east] {-1};
  \filldraw[black] (0,1.5) circle node[anchor=east] {$h^+$};
  
  \draw[->] (7,0) -- (12.5,0);
  \draw[->] (7,-1.5) -- (7,1.5);
  \draw[line width = 3pt] (11,-1.5) -- (11,1.5);
  \draw[line width = 2pt] (7,-1) -- (8.5,-1);
  \draw[dashed] (8.5,-1) -- (8.5,1);
  \draw[line width = 2pt] (8.5,1) -- (9,1);
  \draw[dashed] (9,1) -- (9,-1);
  \draw[line width = 2pt] (9,-1) -- (10.5,-1);
  \draw[dashed] (10.5,-1) -- (10.5,1);
  \draw[line width = 2pt] (10.5,1) -- (11,1);
  \draw[dashed, line width=0.5pt] (7,1) -- (11,1);
  \draw[dashed, line width=0.5pt] (7,-1) -- (11,-1);
  
  \filldraw[black] (7,0) circle node[anchor=east] {0};
  \filldraw[black] (8.5,0) circle node[anchor=north] {1.5};
  \filldraw[black] (9,0) circle node[anchor=north] {2};
  \filldraw[black] (10.5,0) circle node[anchor=north] {3.5};
  \filldraw[black] (11,0) circle node[anchor=north west] {$T=4$};
  \filldraw[black] (12.5,0) circle node[anchor=north west] {$t$};
  \filldraw[black] (7,-1) circle node[anchor=east] {-1};
  \filldraw[black] (7,1) circle node[anchor=east] {+1};
  \filldraw[black] (7,1.5) circle node[anchor=east] {$h^-$};
\end{tikzpicture}
    \caption{An example of function $h$.}
    \label{fig:demonstrationOfh}
\end{figure}

Based on such notations,
 we write down the continuous version of $A_l$ as
\begin{equation}
 	A_{(r^+,r^-)}^{D,\sgn}(t,[\tau_1,\cdots,\tau_D]).
	\label{notationA}
\end{equation}
In the example given in \Cref{fig:demonstrationOfh},
 $A$ can be written as
\begin{equation}
    A_{(+1,-1)}^{5,[+,-,-,+,-]}(t,[1.0,0.5,0.5,0.5,1.0]).
    \label{representationOfAExample}
\end{equation}
An advantage of the notation \eqref{notationA} 
 is that we can define the value of \eqref{notationA} for $(\tau_1,\cdots,\tau_D) \in \partial \triangle_T^{(D)}$
 by taking the limit. 
 In other words,
 the domain of definition for $(\tau_1,\cdots,\tau_D)$ 
 can be extended from $\triangle_T^{(D)}$ to $\overline{\triangle_T^{(D)}}$, 
 which is important for us to close the system
 and formulate the boundary conditions in future sections.
Note that when $(\tau_1, \cdots, \tau_D)$ takes values on $\partial \triangle_T^{(D)}$, the path segment $h$ may have one or more spin flips at the same time, or have spin flips at the initial time or the final time.
Nevertheless, both notations $A(t,h)$ and \eqref{notationA} will be used in our further discussions according to the context.
The precise relation between $A(t,h)$ and $A_{k-\dk}$ will be detailed later in \Cref{ContinuityOfA}.

In i-QuAPI, another quantity
 $\Lambda(S_k,\cdots,S_{k-\dk})$ is also determined by a path segment, while it 
 has one more argument
 than $A_{k-\dk}(S_{k-1},\cdots,S_{k-\dk})$.
In the continuous form, the length of this extra argument becomes infinitesimal so that the associated continuous quantity is also expected to be determined by $h$.
 This will be discussed in detail in \Cref{LimitOfPropagator}. The integral over the paths, which forms the density matrix, will be considered in \Cref{sec:DensityMatrix}.
 
In the following sections,
 our major task is to take the continuous limit $\Delta t \rightarrow 0$
 to define the quantities we need to formulate differential equations.
While taking the limit of the discrete path,
 we assume that the memory length $T = \dk \dt$ is fixed,
 and $S_{k-j}$ takes the value of a fixed path segment $h(\tau)$ with $\tau \in [0,T)$ being the continuous time:
\begin{equation}
    S_{k-j} = h\big((\dk-j)\dt\big), \qquad j = 1,\cdots,\dk.
    \label{hS}
\end{equation}
With this assumption, it is clear that when $\dt$ decreases (or when $\dk$ increases), the function $s_{\dt}$ defined in \eqref{DefSdt} converges to the path segment $h(\tau)$. Thereby, we would like to provide proper definitions for the functions with continuous times.

\subsection{Continuous limit of the propagator}
\label{LimitOfPropagator}
According to \eqref{propagator} and \eqref{definitionI},
 the discrete propagator $\Lambda$ can be expressed by
\begin{equation}
    \begin{split}
	\Lambda(S_k,\cdots,S_{k-\dk}) = 
	&\exp\left(-
	\sum_{m = 0}^{\dk} (s_k^+ - s_k^-) (\eta_{k,k-m} s_{k-m}^+ - \eta_{k,k-m}^* s_{k-m}^-)
	\right)  \\
	&\mel{s_k^+}{\e^{-\ii H_0 \dt}}{s_{k-1}^+}
	\mel{s_{k-1}^-}{\e^{\ii H_0 \dt}}{s_k^-}.
	\end{split}
	\label{discreteLambda}
\end{equation}
In the continuous limit, all the arguments of $\Lambda$ become a path segment $h$, and $\eta_{k,k'}$ needs to be replaced by the bath response function \cite{makri1995tensor1} defined by\footnote{Compared with the expression in \cite{makri1995tensor1}, we have omitted the term with $(s^{+})^2 - (s^{-})^2$ since both $s^+$ and $s^-$ only take values $\pm 1$. Also, the integral domain $[0,+\infty)$ comes from our assumption that all the harmonic oscillators have positive frequencies.}
\begin{equation*}
    \tilde{\eta} (\tau)
    = \frac{1}{\pi} \int_0^\infty \dif \omega
    J(\omega)
    \left(\coth\left(\frac{\beta\omega}{2}\right)\cos(\omega \tau) - \ii \sin(\omega \tau)\right).
\end{equation*}
When $\tau=(k-k')\dt$,
the relation between $\eta_{k,k'}$ and $\tilde{\eta}(\tau)$ is 
\begin{equation}
    \tilde{\eta}(\tau) = \lim_{\dt\rightarrow 0} \frac{\eta_{k,k'}}{\dt^2}.
    \label{eta_teta}
\end{equation}

According to the correspondence between $h(\cdot)$ and $S_j$ given by \eqref{hS}, the continuous limit of $s_{k-1}^\pm$ appearing in the bra-ket notations in \eqref{discreteLambda} should be the left limit of $h^{\pm}(t)$ at $t = T$.
In the case where $h$ is left continuous at time $T$,
 the following proposition holds. 
\begin{proposition}
    \label{thm_propagator}
    Given a path segment $h$, suppose that the discrete path is chosen according to \eqref{hS} and
    $S_k = S_{k-1}$.
    It holds that
    \begin{equation*}
        \lim_{\dt\rightarrow 0}
        \frac{\Lambda(S_k,\cdots,S_{k-\dk})-1}{\dt}
        = -W(h)
    \end{equation*}
    with
    \begin{equation}
    \begin{split}
        W(h) &= \int_0^T \left((h^+(T) - h^-(T)\right)
        \left(\tilde{\eta}(\tau)h^+(T-\tau)
        -\tilde{\eta}^*(\tau)h^-(T-t)\right) \dif \tau \\
        & \quad + \ii \left(
        \mel{h^+(T)}{H_0}{h^+(T)}
		- \mel{h^-(T)}{H_0}{h^-(T)}
        \right).
    \end{split}
    \label{expressionW}
    \end{equation}
    where $(h^+(T),h^-(T)) = (s_{k}^+,s_k^-)$.
    \begin{proof}
        By Taylor expansion of the operators
        $\e^{\pm \ii H_0 \dt} = \id \pm \ii H_0 \dt + \mathcal{O}(\dt^2)$,
        the limit of the bra-kets in \eqref{discreteLambda} can be given by
        \begin{equation*}
            \lim_{\dt\rightarrow 0} 
		    \frac{
		    \mel{s_k^+}{\e^{-\ii H_0\dt}}{s_{k-1}^+}
		     \mel{s_{k-1}^-}{\e^{\ii H_0\dt}}{s_k^-} - 1}{\dt}
		    = -\ii \left(
		    \mel{h^+(T)}{H_0}{h^+(T)}
		    - \mel{h^-(T)}{H_0}{h^-(T)}
	    	\right).
        \end{equation*}
        Here due to the left continuity of $h$, the limits of $s_k^{\pm}$ and $s_{k-1}^{\pm}$ are the same.
        According to the relation of
         discrete $\eta_{k,k'}$ and continuous $\tilde{\eta}$
         in \eqref{eta_teta},
         the following limit holds:
        \begin{equation*}
            \begin{split}
            &\lim_{\dt\rightarrow 0}
    		\frac{\sum_{m = 0}^{\dk} \left(s_k^+ - s_k^-\right) \left(\eta_{k,k-m} s_{k-m}^+ 
    		- \eta_{k,k-m}^* s_{k-m}^-\right)}
    		{\dt} \\
    		=& \int_{0}^T (h^+(T) - h^-(T))
    		\left(\tilde{\eta}(\tau)h^+(T-\tau) - \tilde{\eta}^*(\tau)h^-(T-\tau)\right) \dif \tau.
    		\end{split}
        \end{equation*}
        The proposition follows by combining the two limits
         and applying the Taylor expansion $\e^{x} = 1 + x + \mathcal{O}(x^2)$.
    \end{proof}
\end{proposition}
For a given path $h$ with $(\tau_1,\cdots,\tau_D) \in \triangle_T^{(D)}$,
 one can always increase $\dk$ to satisfy the condition $S_k = S_{k-1}$.
For the case when $S_k\not=S_{k-1}$,
 more details will be discussed in \Cref{sec:boundaryCondition}.

According to this proposition, the path-dependent function $W(h)$ can be regarded as the continuous form of the propagator. As $h$ can be fully determined by the set of parameters described in \Cref{RepresentationOfPaths}, we can mimic the notation in \eqref{notationA} to write $W$ as
\begin{equation*}
    W_{(r^+,r^-)}^{D,\sgn}
    \left([\tau_1,\cdots,\tau_D]\right).
\end{equation*}

\subsection{Continuous limit of $A$'s}
\label{ContinuityOfA}
In this section, we would now like to explore the relation between the function $A(t,h)$ and the quantities $A_{k-\dk+1}(S_k, \cdots, S_{k-\dk+1})$. By the idea of the i-QuAPI method, the quantity $A_{k-\dk+1}(S_k, \cdots, S_{k-\dk+1})$ denotes the ensemble of all the discrete paths whose last segments of length $T$ are given by $S_{k-\dk+1}, \cdots, S_k$. This can be observed by applying 
\eqref{propagation} recursively, resulting in
\begin{equation}
  \begin{split}
	A_{k-\dk+1} (S_k,\cdots,S_{k-\dk+1})
	= &\sum_{S_{k-\dk},\cdots,S_0 \in\mathcal{S}^2} \Bigg[
	\mel{s_k^+}{\e^{-\ii H_0 \dt}}{s_{k-1}^+}
	\cdots
	\mel{s_1^+}{\e^{-\ii H_0 \dt}}{s_0^+}
	\mel{s_0^+}{\tilde{\rho}_s(0)}{s_0^-}\\
	& \mel{s_0^-}{\e^{\ii H_0\dt}}{s_1^-}
	\cdots
	\mel{s_{k-1}^-}{\e^{\ii H_0\dt}}{s_k^-} \\
	&\prod_{j_1=0}^k \prod_{j_2 = \max\{0, k - \dk\}}^{j_1}
	\exp \left( - (s_{j_1}^+ - s_{j_1}^-)
	 (\eta_{j_1,j_2} s_{j_2}^+ - \eta_{j_1,j_2}^* s_{j_2}^-) \right) \Bigg],
  \end{split} 
  \label{ExpressionA}
\end{equation}
where the summation symbol takes into account all the possible paths before the segment specified in the arguments.

We first take the limit of the system part, represented by the product of several bra-kets in \eqref{ExpressionA}. Since the path segment $h$ is denoted by specifying the locations of spin flips, we choose to deal with the continuous segments of $h$ and the discontinuities in $h$ separately, requiring the following two lemmas:
\begin{lemma}
	\label{nonJumpPart}
	Suppose in the time interval $[\mathcal{T}_1,\mathcal{T}_2] = [k_1\dt,k_2\dt]$,
	 the state of the system remains unchanged, \textit{i.e.}
	 \begin{displaymath}
	 S_{k_1} = \cdots = S_{k_2} = S := (r^+,r^-).
	 \end{displaymath}
	 For fixed $\mathcal{T}_1$ and $\mathcal{T}_2$, it holds that
	\begin{gather*}
		\lim_{\dt\rightarrow 0} 
		\mel{s_{k_2}^+}{\e^{-\ii H_0\dt}}{s_{k_2-1}^+} \cdots
		\mel{s_{k_1+1}^+}{\e^{-\ii H_0\dt}}{s_{k_1}^+}
		= \e^{-\ii 
		\expval{H_0}{r^+} (\mathcal{T}_2-\mathcal{T}_1)}, \\
		\lim_{\dt\rightarrow 0} 
		\mel{s_{k_1}^-}{\e^{\ii H_0\dt}}{s_{k_1+1}^-} \cdots
		\mel{s_{k_2-1}^-}{\e^{\ii H_0\dt}}{s_{k_2}^-}
		= \e^{\ii \expval{H_0}{r^-} (\mathcal{T}_2-\mathcal{T}_1)}.
		\label{tempLable1}
	\end{gather*}
	\begin{proof}
		By the limit 
		$\e = \lim_{x\rightarrow 0}\left(1+x\right)^{1/x}$
		and Taylor expansion of the operator 
		$\e^{-\ii H_0\dt} = \id - \ii H_0 \dt + \mathcal{O}(\dt^2)$,
		\begin{equation*}
		\begin{split}
			&\lim_{\dt\rightarrow 0}
			 \mel{s_{k_2}^+}{\e^{-\ii H_0\dt}}{s_{k_2-1}^+} \cdots
		    \mel{s_{k_1+1}^+}{\e^{-\ii H_0\dt}}{s_{k_1}^+} \\
			=& \lim_{\dt\rightarrow 0}
			\expval{\id-\ii H_0 \dt + \mathcal{O}(\dt^2)}{r^+}
			^{(\mathcal{T}_2-\mathcal{T}_1)/\dt} \\
			=& \lim_{\dt\rightarrow 0}
			\left( 1- \ii\dt \expval{H_0}{r^+} + \mathcal{O}(\dt^2) \right)
			^{(\mathcal{T}_2-\mathcal{T}_1)/\dt} \\
			=&\e^{-\ii \expval{H_0}{r^+}
			 (\mathcal{T}_2-\mathcal{T}_1)}.
		\end{split}
		\end{equation*}
		The proof for the second limit is similar.
	\end{proof}
\end{lemma}
\begin{lemma}
	\label{JumpTheorem}
	When $s_k^+ \not= s_{k+1}^+$, the following limit holds.
	\begin{equation}
		\lim_{\dt\rightarrow 0}
		\frac{\mel{s_{k+1}^+}{\e^{-\ii H_0\dt}}{s_k^+}}{\dt}
		= -\ii \mel{s_{k+1}^+}{H_0}{s_k^+}.
	\end{equation}
	Similarly, when $s_k^- \not= s_k^-$, we have
	\begin{equation}
		\lim_{\dt\rightarrow 0}
		\frac{\mel{s_k^-}{\e^{\ii H_0 \dt}}{s_{k+1}^-}}{\dt}
		= \ii \mel{s_k^-}{H_0}{s_{k+1}^-}.
		\label{tempLabel2}
	\end{equation}
	\begin{proof}
		By Taylor expansion of $\e^{-\ii H_0\dt}$,
		\begin{align*}
			\lim_{\dt\rightarrow 0}
			\frac{\mel{s_{k+1}^+}{\e^{-\ii H_0 \dt}}{s_k^+}}{\dt}
			&=\lim_{\dt\rightarrow 0}
			\frac{\mel{s_{k+1}^+}{\id-\ii H_0 \dt + \mathcal{O}(\dt^2)}{s_k^+}}
			{\dt} \\
			&=\lim_{\dt\rightarrow 0}
			\frac{-\ii\dt\mel{s_{k+1}^+}{H_0}{s_k^+}+\mathcal{O}(\dt^2)}{\dt}
			= -\ii \mel{s_{k+1}^+}{H_0}{s_k^+}.
		\end{align*}
		The proof for (\ref{tempLabel2}) is similar.
	\end{proof}
\end{lemma}

This lemma shows that the quantity $A_{k-\dk+1}(S_k,\cdots,S_{k-\dk+1})$
 has magnitude $\dt^{D}$ 
 if $s_k^{\pm}$ changes values $D$ times.
Thus, it is natural to let the function $A(t,h)$ demonstrated in \Cref{RepresentationOfPaths} be the limit of $\dt^{-D} A_{k-\dk+1}(S_k,\cdots,S_{k-\dk+1})$. 
We write down the result of such a limit in the following definition:
\begin{definition}
    \label{definitionACont}
    For a path $h$ given by \eqref{contPath},
    the quantity $A(t,h)$ or equivalently, $A_{(r_0^+,r_0^-)}^{D,\sgn}(t,[\tau_1,\cdots,\tau_D])$
     is defined by
    \begin{equation}
    A(t,h) = \sum_{\tilde{D}=0}^\infty
        \sum_{(\tilde{r}_0^+,\tilde{r}_0^-)\in\mathcal{S}^2}
        \sum_{ \widetilde{\sgn} \in \{+,-\}^{\tilde{D}} }
        \int_{\triangle_t^{(\tilde{D})}} \dif \tilde{\boldsymbol{\tau}}
        X(g)
        \delta_{\tilde{r}_{\tilde{D}}^+}^{r_0^+}
        \delta_{\tilde{r}_{\tilde{D}}^-}^{r_0^-}
        Y(h)
        \e^{Z(g,h)}.
    \label{Asimplified}
    \end{equation}
    with
    \begin{gather}
    \begin{split}
        X(g)=&\mel{\tilde{r}_0^+}{\rho_s(0)}{\tilde{r}_0^-}
         \e^{-\ii\left(\mel{\tilde{r}_0^+}{H_0}{\tilde{r}_0^+}-\mel{\tilde{r}_0^-}{H_0}{\tilde{r}_0^-}\right)\tilde{\tau}_1} \\
        &\prod_{j=1}^{\tilde{D}-1}
        \Bigg(
        \e^{-\ii\big(\mel{\tilde{r}_j^+}{H_0}{\tilde{r}_j^+}-\mel{\tilde{r}_j^-}{H_0}{\tilde{r}_j^-}\big)\tilde{\tau}_{j+1}}
        \left(-\ii\mel{\tilde{r}_j^+}{H_0}{\tilde{r}_{j-1}^+}\delta_{\widetilde{\sgn}_j}^+ + \ii \mel{\tilde{r}_{j-1}^-}{H_0}{\tilde{r}_j^-}\delta_{\widetilde{\sgn}_j}^-\right)
        \Bigg) \\
        &\e^{-\ii\big(\mel{\tilde{r}_{\tilde{D}}^+}{H_0}{\tilde{r}_{\tilde{D}}^+}-\mel{\tilde{r}_{\tilde{D}}^-}{H_0}{\tilde{r}_{\tilde{D}}^-}\big)\left(t-\sum_{j=1}^{\tilde{D}}\tilde{\tau}_j\right)}
        \left(-\ii\mel{\tilde{r}_{\tilde{D}}^+}{H_0}{\tilde{r}_{{\tilde{D}}-1}^+}\delta_{\widetilde{\sgn}_{\tilde{D}}}^+ + \ii \mel{\tilde{r}_{{\tilde{D}}-1}^-}{H_0}{\tilde{r}_{\tilde{D}}^-}\delta_{\widetilde{\sgn}_{\tilde{D}}}^-\right);
    \end{split} \label{Xeq} \\
    \begin{split}
        Y(h) = & \e^{-\ii\left(\mel{{r}_0^+}{H_0}{{r}_0^+}-\mel{{r}_0^-}{H_0}{{r}_0^-}\right){\tau}_1}
        \\
        &\prod_{j=1}^{D-1}
        \Bigg(
        \e^{-\ii\big(\mel{r_j^+}{H_0}{r_j^+}-\mel{r_j^-}{H_0}{r_j^-}\big)\tau_{j+1}}
        \left(-\ii\mel{r_j^+}{H_0}{r_{j-1}^+}\delta_{\sgn_j}^+ + \ii \mel{r_{j-1}^-}{H_0}{r_j^-}\delta_{\sgn_j}^-\right)
        \Bigg) \\
        &\e^{-\ii\big(\mel{r_D^+}{H_0}{r_D^+}-\mel{r_D^-}{H_0}{r_D^-}\big)\left(T-\sum_{j=1}^D \tau_j\right)}
        \left(-\ii\mel{r_D^+}{H_0}{r_{D-1}^+}\delta_{\sgn_j}^+ + \ii \mel{r_{D-1}^-}{H_0}{r_D^-}\delta_{\sgn_j}^-\right);
    \end{split} \label{Yeq} \\
    \begin{split}
        Z(g,h) =& -\int_0^t\dif x_1 \int_{\max\{0,x_1-T\}}^{x_1}\dif x_2
        (g^+(x_1)-g^-(x_1))
        (g^+(x_2)\tilde{\eta}(x_1-x_2)-g^-(x_2)\tilde{\eta}^*(x_1-x_2)) \\
        & -\int_t^{t+T}
        \dif x_1 \int_{\max\{0,x_1-T\}}^t\dif x_2
        (h^+(x_1-t)-h^-(x_1-t))
        (g^+(x_2)\tilde{\eta}(x_1-x_2)-g^-(x_2)\tilde{\eta}^*(x_1-x_2)) \\
        & -\int_{0}^T \dif x_1 \int_0^{x_1} \dif x_2
        (h^+(x_1)-h^-(x_1))
        \left(h^+(x_2)\tilde{\eta}(x_1-x_2)-h^-(x_2)\tilde{\eta}^*(x_1-x_2)\right)
    \end{split} \label{Zeq}
\end{gather}
and $g:[0,t)\rightarrow \mathcal{S}^2$ is defined by
\begin{equation}
    g(\tau) =\begin{cases}
    (\tilde{r}_0^+,\tilde{r}_0^-), & \forall\tau\in[0,\tilde{\tau}_1) \\
    (\tilde{r}_k^+,\tilde{r}_k^-), &
    \forall \tau\in[\sum_{j=1}^k \tilde{\tau}_j,\sum_{j=1}^{k+1}\tilde{\tau}_j),
    \quad k = 1,\cdots,\tilde{D-1} \\
    (\tilde{r}_{\tilde{D}}^+,\tilde{r}_{\tilde{D}}^-), &
    \forall \tau\in[\sum_{j=1}^{\tilde{D}}\tilde{\tau}_j,t)
    \end{cases}
    \label{function_g}
\end{equation}
with $\boldsymbol{\tau}=(\tilde{\tau}_1,\cdots,\tilde{\tau}_D)
\in \triangle_t^{(\tilde{D})}$.
In \eqref{Asimplified},
 $\delta_{\tilde{r}_{\tilde{D}}^+}^{r_0^+}=\begin{cases}
 1, & \text{if } \tilde{r}_{\tilde{D}}^+ = {r_0^+} \\
 0, & \text{if } \tilde{r}_{\tilde{D}}^+ \not= {r_0^+}
 \end{cases}$
and
$\delta_{\tilde{r}_{\tilde{D}}^-}^{r_0^-}=\begin{cases}
 1, & \text{if } \tilde{r}_{\tilde{D}}^- = {r_0^-} \\
 0, & \text{if } \tilde{r}_{\tilde{D}}^- \not= {r_0^-}
 \end{cases}$.
In \eqref{Xeq} and \eqref{Yeq},
 $\sgn_k$ represents the $k$-th component of $\sgn$, 
 $\delta_{\sgn_j}^{+}=\begin{cases}
 1, & \text{if } \sgn_j \text{ is } + \\
 0, &  \text{if } \sgn_j \text{ is } -
 \end{cases}$
and
 $\delta_{\sgn_j}^{-}=\begin{cases}
 1, & \text{if } \sgn_j \text{ is } - \\
 0, &  \text{if } \sgn_j \text{ is } +
 \end{cases}$.
\end{definition}
One can observe from the above definition
 that the path is separated into two segments. 
The first segment, denoted by $g$,
 covers the interval $[0,t)$,
 and all possibilities of $g$ have been considered
 and summed up on the right-hand side of \eqref{Asimplified}.
The second segment, denoted by $h$, 
 is specified in the arguments of $A$. 
The two delta symbols in \eqref{Asimplified} indicate
 that the last state of $g$ agrees with the first state of $h$, so that the two segments are connected without discontinuities.
The contributions of $g$ and $h$ coming from the system evolution are summarized in $X(g)$ and $Y(h)$, respectively,
 and the non-Markovian contribution coming from the system-bath interaction is provided in the term $\e^{Z(g,h)}$, which couples both segments of the path.
Note that when $t=0$, 
 the value of $A(0,h)$ is not necessarily zero.
The only terms contributing to the final result are the terms with $\tilde{D}=0$
 while other terms vanish as the integral domain $\triangle_t^{(D)}$ has measure zero. 
The formulation of $A(t,h)$ follows the same idea of its discrete counterpart $A_{k-\dk}(S_{k-1},\cdots,S_{k-\dk})$,
 and the relation between these two quantities are given in the following theorem:
\begin{theorem}
    \label{Acontanddis}
    For any $t \geqslant 0$ and given $h$,
    the following limit holds:
    \begin{equation*}
        A(t,h) = \lim_{\dt\rightarrow 0}
        \frac{A_{k-\dk}(S_{k-1},\cdots,S_{k-\dk})}{\dt^D}
    \end{equation*}
    where $S_{k-\dk},\cdots,S_{k-1}$ is given by \eqref{hS}.
    \begin{proof}
        For any given $\tilde{D}$, $(r_0^+,r_0^-)\in\mathcal{S}^2$
        and $\widetilde{\sgn}\in\{+,-\}^{\tilde{D}}$,
        we first estimate $X(g), Y(h)$ and $Z(g,h)$. Let $C_1 = \max_{r,s \in \mathcal{S}} |\matrixel{r}{H_0}{s}|$. Then
        \begin{equation*}
            \vert X(g) \vert \leqslant 
            C_1^{\tilde{D}}, \qquad
            \vert Y(h) \vert \leqslant
            C_1^{D}.
        \end{equation*}
        In addition, since $\tilde{\eta}(\cdot)$ is a continuous function on $[0,T]$, we have
        \begin{equation*}
            \vert Z(g,h) \vert \leqslant
            \frac{T^2+2Tt}{2}\cdot 4\max\vert\tilde{\eta}\vert
            :=C_2(t).
        \end{equation*}
        Therefore, we have
        \begin{equation}
            \begin{split}
                R_1:=&\left\vert\sum_{\tilde{D}=\overline{D}}^\infty
                \sum_{(\tilde{r}_0^+,\tilde{r}_0^-)\in\mathcal{S}^2}
                \sum_{ \widetilde{\sgn} \in \{+,-\}^{\tilde{D}} }
                \int_{\triangle_t^{\tilde{D}}}\dif \tilde{\boldsymbol{\tau}}
                X(g)
                \delta_{\tilde{r}_{\tilde{D}}^+}^{r_0^+}
                \delta_{\tilde{r}_{\tilde{D}}^-}^{r_0^-}
                Y(h)
                \e^{Z(g,h)}\right\vert \\
                \leqslant &\sum_{\tilde{D}=\overline{D}}^\infty
                \sum_{(\tilde{r}_0^+,\tilde{r}_0^-)\in\mathcal{S}^2}
                \sum_{ \widetilde{\sgn} \in \{+,-\}^{\tilde{D}} }
                \int_{\triangle_t^{\tilde{D}}}\dif \tilde{\boldsymbol{\tau}}
                C_1^{\tilde{D}+D}C_2(t) \\
                \leqslant& \sum_{\tilde{D}=\overline{D}}^\infty
                4\cdot 2^{\tilde{D}} C_1^{\tilde{D}+D} C_2(t) \frac{t^{\tilde{D}}}{\tilde{D}!}
                \rightarrow 0
                \quad\text{as}\quad \overline{D}\rightarrow\infty.
            \end{split}
            \label{R1}
        \end{equation}
        Similarly, in \eqref{ExpressionA},
         for given $h$, $S_{k-\dk+1},\cdots,S_k$ 
         are determined by $h$ as in \eqref{hS}.
        \begin{equation*}
        \begin{split}
        	R_2:={}&\Bigg\vert\sum_{\substack{S_{k-\dk},\cdots,S_0 \in\mathcal{S}^2\\ \text{number of spin flips}\geqslant \overline{D}}}
        	\mel{s_k^+}{\e^{-\ii H_0 \dt}}{s_{k-1}^+}
        	\cdots
        	\mel{s_1^+}{\e^{-\ii H_0 \dt}}{s_0^+}
        	\mel{s_0^+}{\tilde{\rho}_s(0)}{s_0^-}\\
        	&\qquad \mel{s_0^-}{\e^{\ii H_0\dt}}{s_1^-}
        	\cdots
        	\mel{s_{k-1}^-}{\e^{\ii H_0\dt}}{s_k^-} 
        	\prod_{j_1=0}^k \prod_{j_2 = \max\{0, k - \dk\}}^{j_1}
        	\exp \left( - (s_{j_1}^+ - s_{j_1}^-)
        	 (\eta_{j_1,j_2} s_{j_2}^+ - \eta_{j_1,j_2}^* s_{j_2}^-) \right) \Bigg\vert \\
        	\leqslant{} &
        	\sum_{\tilde{D} = \overline{D}}^{2\dk-2}
        	\sum_{\substack{S_{k-\dk},\cdots,S_0 \in\mathcal{S}^2\\ \text{number of spin flips} =  \overline{D}}} 
        	\Bigg\vert
        	\mel{s_k^+}{\e^{-\ii H_0 \dt}}{s_{k-1}^+}
        	\cdots
        	\mel{s_1^+}{\e^{-\ii H_0 \dt}}{s_0^+}
        	\mel{s_0^+}{\tilde{\rho}_s(0)}{s_0^-}\\
        	&\qquad \mel{s_0^-}{\e^{\ii H_0\dt}}{s_1^-}
        	\cdots
        	\mel{s_{k-1}^-}{\e^{\ii H_0\dt}}{s_k^-}
        	\prod_{j_1=0}^k \prod_{j_2 = \max\{0, k - \dk\}}^{j_1}
        	\exp \left( - (s_{j_1}^+ - s_{j_1}^-)
        	 (\eta_{j_1,j_2} s_{j_2}^+ - \eta_{j_1,j_2}^* s_{j_2}^-) \right) \Bigg\vert \\
        	 \lesssim & \sum_{\tilde{D}=\overline{D}}^{2\dk-2}
        	 {{2\dk-2}\choose{\tilde{D}}}
        	 \dt^{\tilde{D}+D}
        	 \leqslant \sum_{\tilde{D}=\overline{D}}^{\infty}
        	 \frac{(2\dk-2)^{\tilde{D}}}{\tilde{D}!} \dt^{\tilde{D}+D}
        	 \leqslant \sum_{\tilde{D}=\overline{D}}^\infty \frac{(2T)^{\tilde{D}}}{\tilde{D}!}\dt^D.
          \end{split}
        \end{equation*}
        Therefore, $\dt^{-D}R_2 \rightarrow 0$ as $\overline{D}\rightarrow\infty$.
        For any $\varepsilon>0$,
         there exist a sufficiently large integer $\overline{D}$, such that
         \begin{equation*}
         R_1 < \varepsilon \quad \text{and} \quad \dt^{-D}R_2 < \varepsilon.
         \end{equation*}
        For the integral over $\triangle_t^{\tilde{D}}$,
         according to definition of Riemann integral,
         for a large enough $\dk$,
        \begin{equation*}
            \left\vert
            \int_{\triangle_t^{\tilde{D}}} 
            \dif \tilde{\boldsymbol{\tau}}
            X(g)
            \delta_{\tilde{r}_{\tilde{D}}^+}^{r_0^+}
            \delta_{\tilde{r}_{\tilde{D}}^-}^{r_0^-}
            Y(h)
            \e^{Z(g,h)}
            - \sum_{g\in\mathcal{G}_{(\tilde{r}_0^+,\tilde{r}_0^-)}^{\tilde{D}}} X(g)\delta_{\tilde{r}_{\tilde{D}}^+}^{r_0^+}
            \delta_{\tilde{r}_{\tilde{D}}^-}^{r_0^-}
            Y(h)
            \e^{Z(g,h)} \dt^{\tilde{D}} \right\vert
            < \frac{\varepsilon}{\tilde{D}!}
        \end{equation*}
        with $\mathcal{G}_{(\tilde{r}_0^+,\tilde{r}_0^-)}^{\tilde{D}}$
        being the set of paths 
        that have initial states $(\tilde{r}_0^+,\tilde{r}_0^-)$ and $\tilde{D}$ spin flips
        in the set of all possible time discrete paths $S_{k-\dk},\cdots,S_{k-1}$.
        
        In addition, for any given path $g\in\mathcal{G}_{(\tilde{r}_0^+,\tilde{r}_0^-)}^{\tilde{D}}$
         with $\tilde{D}$ spin flips,
         according to \Cref{nonJumpPart}, \Cref{JumpTheorem}, \eqref{eta_teta} and the definition of Riemann integral,
        there exists a large enough $\dk$, such that 
        \begin{equation*}
        \begin{split}
            \Bigg\vert & \dt^{\tilde{D}}
                X(g) \delta_{\tilde{r}_{\tilde{D}}^+}^{r_0^+}
                \delta_{\tilde{r}_{\tilde{D}}^-}^{r_0^-}
                Y(h)
                \e^{Z(g,h)} - 
                \dt^{-D}
            \mel{s_k^+}{\e^{-\ii H_0 \dt}}{s_{k-1}^+}
        	\cdots
        	\mel{s_1^+}{\e^{-\ii H_0 \dt}}{s_0^+}
        	\mel{s_0^+}{\tilde{\rho}_s(0)}{s_0^-}\\
        	& \mel{s_0^-}{\e^{\ii H_0\dt}}{s_1^-}
        	\cdots
        	\mel{s_{k-1}^-}{\e^{\ii H_0\dt}}{s_k^-} 
        	\prod_{j_1=0}^k \prod_{j_2 = \max\{0, k - \dk\}}^{j_1}
        	\exp \left( - (s_{j_1}^+ - s_{j_1}^-)
        	 (\eta_{j_1,j_2} s_{j_2}^+ - \eta_{j_1,j_2}^* s_{j_2}^-) \right) \Bigg\vert < \varepsilon\dt^{\tilde{D}}
        \end{split}
        \end{equation*}
        where $S_j = g(j\dt)$ for $j = 0,\cdots, k-\dk-1$.
        
        Combine all the results,
         we have
        \begin{equation*}
        \begin{split}
            &\left\vert A(t,h) - \frac{A_{k-\dk}(S_{k-1},\cdots,S_{k-\dk})}{\dt^D}
            \right\vert\\
            \leqslant& R_1 + \dt^{-D}R_2 
            + \sum_{\tilde{D}=0}^{\overline{D}-1}
            4\cdot 2^{\tilde{D}}
            \left(\frac{\varepsilon}{\tilde{D}!}
            + {{2\dk-2}\choose{\tilde{D}}} \varepsilon\dt^{\tilde{D}}
            \right) \\
            < & \left(2+4\e^2 + 4\e^{4T}\right)\varepsilon
        \end{split}
        \end{equation*}
        where $4\cdot 2^{\tilde{D}}$ comes from the summation over $(\tilde{r}_0^+,\tilde{r}_0^-)$ and $\widetilde{\sgn}$,
        the binomial coefficient is the cardinality of $\mathcal{G}_{(\tilde{r}_0^+,\tilde{r}_0^-)}^{\tilde{D}}$.
        The proof is completed as $\varepsilon$ is arbitrary.
    \end{proof}
\end{theorem}
According to the definition of $A(t,h)$ for  $(\tau_1,\cdots,\tau_D)\in\triangle_T^{(D)}$ in \Cref{definitionACont},
 we can extend the definition of $A(t,h)$
 to $(\tau_1,\cdots,\tau_D)\in\partial\triangle_T^{(D)}$
 by taking the limit.
The following theorem guarantees that such extension is well defined.
\begin{theorem}
\label{boundaryADefn}
    For $\boldsymbol{\tau}' = (\tau_1',\cdots,\tau_D')\in\partial\triangle_T^{(D)}$,
     the limit
    \begin{equation*}
        \lim_{(\tau_1,\cdots,\tau_D)\rightarrow(\tau_1',\cdots,\tau_D')} A_{(r^+,r^-)}^{D,\sgn}(t,[\tau_1,\cdots,\tau_D])
    \end{equation*}
    exists and therefore, $A_{(r^+,r^-)}^{D,\sgn}(t,[\tau_1',\cdots,\tau_D'])$
    can be defined by this limit.
    \begin{proof}
        In order to prove the limit exists,
        we only need to show that
        for any $\varepsilon>0$, 
         there exist a small enough positive number $\delta$,
         such that
        \begin{equation*}
            \left\vert
            A_{(r^+,r^-)}^{D,\sgn}(t,[\tau_1^{(1)},\cdots,\tau_D^{(1)}]) - 
            A_{(r^+,r^-)}^{D,\sgn}(t,[\tau_1^{(2)},\cdots,\tau_D^{(2)}]) \right\vert < \varepsilon
        \end{equation*}
        for any $\boldsymbol{\tau}_1 = (\tau_1^{(1)},\cdots,\tau_D^{(1)})$ and 
        $\boldsymbol{\tau}_2 = (\tau_1^{(2)},\cdots,\tau_D^{(2)})$
        with 
        $\boldsymbol{\tau}_1,\boldsymbol{\tau}_2 \in \triangle_{T}^{(D)}$,
        $\Vert\boldsymbol{\tau}_1-\boldsymbol{\tau}'\Vert_1<\delta/2$ and $\Vert\boldsymbol{\tau}_2-\boldsymbol{\tau}'\Vert_1<\delta/2$.
        Denote the path of $A_{(r^+,r^-)}^{D,\sgn}(t,[\tau_1^{(1)},\cdots,\tau_D^{(1)}])$ by $h_1$ and the path of
        $A_{(r^+,r^-)}^{D,\sgn}(t,[\tau_1^{(2)},\cdots,\tau_D^{(2)}])$ by $h_2$.
        $h_1$ and $h_2$ differs only on intervals whose sum is smaller than $\delta$.
        Therefore,
        \begin{equation*}
            \vert Y(h_1) - Y(h_2) \vert
            \leqslant D L_1 C_1^{D} \delta
        \end{equation*}
        where $L_1$ is the Lipsichitz constant such that
        \begin{equation*}
            \left\vert \e^{-\ii\left(\mel{r^+}{H_0}{r^+}-\mel{r^-}{H_0}{r^-}\right)t_1} 
            -\e^{-\ii\left(\mel{r^+}{H_0}{r^+}-\mel{r^-}{H_0}{r^-}\right)t_2}
            \right\vert
            \leqslant L_1 \vert t_1 - t_2 \vert
        \end{equation*}
        for any $(r^+,r^-)\in\mathcal{S}^2$.
        In addition, for any path $g:[0,t) \rightarrow \mathcal{S}^2$,
        \begin{equation*}
            \vert Z(g,h_1) - Z(g,h_2) \vert
            \leqslant 4\delta T\cdot 2\max\vert \tilde{\eta}\vert.
        \end{equation*}
        Then
        \begin{equation*}
            \left\vert \e^{Z(g,h_1)} - \e^{Z(g,h_2)} \right\vert \leqslant 8TL_2\max\vert\tilde{\eta}\vert\delta
        \end{equation*}
        with $L_2$ is the Lipsichitz constant of function $\e^x$ in the interval $\vert x \vert \leqslant C_2(t)$ where $C_2(t)$ is defined in \Cref{Acontanddis}.
        Therefore,
        \begin{equation*}
            \vert X(g)Y(h_1)Z(g,h_1) - X(g)Y(h_2)Z(g,h_2) \vert
            < C_1^{\tilde{D}}C_1^DC_2(t) 
            \left(D L_1C_1^D+8TL_2 \max\vert\tilde{\eta}\vert\right)\delta
            :=C_1^{\tilde{D}}C_3\delta.
        \end{equation*}
        Consequentially we have
        \begin{equation*}
            \vert A(t,h_1) - A(t,h_2) \vert
            < \sum_{\tilde{D}=0}^{\infty}
            \sum_{(\tilde{r}_0^+,\tilde{r}_0^-)\in\mathcal{S}^2}
            \sum_{\widetilde{\sgn}\in\{+,-\}^{\tilde{D}}}
            \int_{\triangle_t^{(\tilde{D})}} \dif \tilde{\boldsymbol{\tau}}
            C_1^{\tilde{D}}C_3\delta
            \leqslant \sum_{\tilde{D}=0}^{\infty}
            4\frac{(2tC_1)^{\tilde{D}}}{\tilde{D}!}C_3\delta
            = 4\e^{2tC_1}C_3\delta.
        \end{equation*}
        The proof is completed by choosing $\delta = (4\e^{2tC_1}C_3)^{-1}\varepsilon$.
    \end{proof}
\end{theorem}
  
\subsection{Density Matrix and Observable} \label{sec:DensityMatrix}
Based on the definition of $A(t,h)$, we would like to represent the density matrix \eqref{densityMatrixDiscrete} using the form with continuous time. On the left-hand side of \eqref{densityMatrixDiscrete}, the state $S_N = (s_N^+, s_N^-)$ denotes the ``final state'' of the path segment. Such a state corresponds to the left limit of $h(\tau)$ at $\tau = T$. For the spin-boson model, it is fully determined by the initial state $(r^+, r^-)$, the number of spin flips $D$ and the sign vector $\sgn$. We denote this final state by
\begin{equation}
    r_f^{\pm} = r_f^{\pm}(r^+, r^-, D, \sgn).
    \label{finalState}
\end{equation}
For example, in \Cref{fig:demonstrationOfh},
 the path segment is represented by the parameters in \eqref{representationOfAExample}
 which determines the final state to be
 $(r_f^+,r_f^-) = (+1,+1)$.

Similar to the discrete case, 
 the values of continuous $A$'s are closely related to the density matrix,
 which contains all information of a quantum system.
Here is the method to construct 
 the reduced density matrix $\tilde{\rho}(t)$
 based on the values of $A$'s.
\begin{theorem}
    \label{densityMatrix}
    The density matrix of a spin-boson model is
	\begin{equation}
		\rho_s (t) = 
		\sum_{D=0}^\infty
		\sum_{(r^+, r^-)\in\mathcal{S}^2} 
		\sum_{\sgn\in\{+,-\}^D}
		\int_{\triangle_T^{(D)}}
		 A_{(r^+,r^-)}^{D,\sgn}
		 \left(t,[\tau_1,\cdots,\tau_D]\right) 
		 \dyad{r_f^+}{r_f^-}
		 \dif \boldsymbol{\tau}
		 \label{densityMatrixContinuous}
	\end{equation}
	with $\triangle_T^{(D)}$ is defined by \eqref{DefnSimplex},
	 $r_f^{\pm}$ is determined by \eqref{finalState}
	 and $\boldsymbol{\tau} = (\tau_1,\cdots,\tau_D)\in\triangle_T^{(D)}$.
    \begin{proof}
        For any fixed $t$
        and a path $h$ with $D$ spin flips,
         by choosing $\overline{D} = 0$ in \eqref{R1}, we have
        \begin{equation*}
             \left\vert A_{(r^+,r^-)}^{D,\sgn}(t,[\tau_1,\cdots,\tau_D])
             \right\vert \leqslant 4 \e^{2 C_1 t} C_1^D C_2(t).
        \end{equation*}
        For any $\overline{D}$, we have
        \begin{equation*}
        \begin{split}
            R_1 :=&\left\Vert\sum_{D = \overline{D}}^{\infty} \sum_{(r^+,r^-)\in\mathcal{S}^2}
            \sum_{\sgn\in\{+,-\}^{D}}
            \int_{\triangle_T^{(D)}}
            A_{(r^+,r^-)}^{D,\sgn}(t,[\tau_1,\cdots,\tau_D])
            \dyad{r_f^+}{r_f^-}
            \dif\boldsymbol{\tau}\right\Vert \\
            \leqslant & \sum_{D = \overline{D}}^{\infty} \sum_{(r^+,r^-)\in\mathcal{S}^2}
            \sum_{\sgn\in\{+,-\}^{D}} 
            \int_{\triangle_T^{(D)}}
            \left\vert A_{(r^+,r^-)}^{D,\sgn}(t,[\tau_1,\cdots,\tau_D])\right\vert \dif\boldsymbol{\tau} \\
            \leqslant & \sum_{D = \overline{D}}^{\infty} \sum_{(r^+,r^-)\in\mathcal{S}^2}
            \sum_{\sgn\in\{+,-\}^{D}}
            \int_{\triangle_T^{(D)}}
            4\e^{2C_1 t} C_1^D C_2(t)
            \dif\boldsymbol{\tau} \\
            =& \sum_{D=\overline{D}}^\infty 
            4\cdot 2^D \frac{T^D}{D!} 4\e^{2C_1 t}C_1^D C_2(t)
            = \sum_{D=\overline{D}}^{\infty}
            16\e^{2C_1 t} C_2(t) \frac{(2C_1 T)^D}{D!}
            \rightarrow 0\quad\text{as}\quad \overline{D}\rightarrow\infty
        \end{split}
        \end{equation*}
        Similarly, for the discrete expression of density matrix, 
         we have the following estimation.
        \begin{equation*}
        \begin{split}
            R_2 := &\left\Vert
            \sum_{\substack{S_{k-\dk+1,\cdots,S_k} \\ \text{number of spin flips}\geqslant\overline{D}}}
            A_{k-\dk+1}(S_k,\cdots, S_{k-\dk+1})
            \dyad{s_k^+}{s_k^-}
            \right\Vert \\
            \lesssim &
            4 \sum_{D=\overline{D}}^{2\dk-2}
            {{2\dk-2}\choose{D}}
            4\e^{2C_1 t}C_1^DC_2(t)\dt^D\\
            \leqslant& 4 \sum_{D=\overline{D}}^{2\dk-2}
            \frac{(2\dk-2)^D}{D!}
            4\e^{2C_1 t}C_1^DC_2(t)\dt^D \\
            \leqslant& \sum_{D=\overline{D}}^\infty 
            16\e^{2C_1 t}C_2(t)\frac{(2C_1T)^D}{D!}
            \rightarrow 0
            \quad\text{as}\quad \overline{D}\rightarrow\infty.
        \end{split}
        \end{equation*}
        
    Therefore, for any $\varepsilon>0$,
     there exists a large enough integer $\overline{D}$, 
     such that $R_1 < \varepsilon, R_2 < \varepsilon$.
    
    For each $D=0,\cdots,\overline{D}$, $(r^+,r^-)\in\mathcal{S}^2$
    and $\sgn\in\{+,-\}^D$, $\dyad{r_f^+}{r_f^-}$ is fixed.
    Therefore, according to the definition of Riemann integral,
    there exists a large enough integer $M_1$ with $M_1>\overline{D}$,
     such that for all and $\dk>M_1$, $D=0,\cdots,\overline{D}$, $(r^+,r^-)\in\mathcal{S}^2$
    and $\sgn\in\{+,-\}^D$,
    \begin{equation*}
        \left\vert 
        \int_{\triangle_T^{(D)}} A_{(r^+,r^-)}^{D,\sgn}(t,[\tau_1,\cdots,\tau_D])
        \dif \boldsymbol{\tau}
        - \sum_{s_{\dt}\in\mathcal{I}_D^{\dk}}
        A(t,s_{\dt})\dt^D
        \right\vert < \frac{\varepsilon}{D!}
    \end{equation*}
    where $\mathcal{I}_D^{\dk}$ is the set of all piecewise constant path segment with $D$ spin flips of continuous time determined by the sequence $S_k,\cdots,S_{k-\dk+1}$.
    According to \Cref{Acontanddis},
     there exists a large enough integer $M_2$, such that
     for $\dk > M_2$
    \begin{equation*}
        \left\vert 
        A(t,s_{\dt})\dt^D
        - A_{k-\dk+1}(S_k,\cdots,S_{k-\dk+1})
        \right\vert < \varepsilon\dt^D
    \end{equation*}
    holds for all $D=0,\cdots,\overline{D}$, $(r^+,r^-)\in\mathcal{S}^2$ and $\sgn\in\{+,-\}^D$.
    For the paths with number of spin flips being $D$ (not greater than $\overline{D}$),
    we have
    \begin{equation*}
    \begin{split}
        &\left\vert\int_{\triangle_T^{(D)}} A_{(r^+,r^-)}^{D,\sgn}(t,[\tau_1,\cdots,\tau_D])
        \dif \boldsymbol{\tau}
        - \sum_{\substack{S_{k-1},\cdots,S_{k-\dk+1}\in\mathcal{S}^2\\ \text{number of spin flips} = D}} A_{k-\dk+1}(S_k,\cdots,S_{k-\dk+1})\right\vert \\
        <& {{2\dk-4}\choose{D}}\varepsilon\dt^D + \frac{\varepsilon}{D!}
        \leqslant \left(\frac{(2\dk-4)^D}{D!}\dt^D + \frac{1}{D!}\right)\varepsilon
        < \frac{(2T)^D+1}{D!}\varepsilon.
    \end{split}
    \end{equation*}
    Therefore, choose $\dk>\max\{M_1,M_2\}$,
     we finally have
    \begin{align*}
        &\Bigg\Vert \sum_{D=0}^\infty
		\sum_{(r^+, r^-)\in\mathcal{S}^2} 
		\sum_{\sgn\in\{+,-\}^D}
		\int_{\triangle_T^{(D)}}
		 A_{(r^+,r^-)}^{D,\sgn}
		 \left(t,[\tau_1,\cdots,\tau_D]\right) 
		 \dyad{r_f^+}{r_f^-}
		 \dif \boldsymbol{\tau}  \\
		 &\quad - \sum_{S_{N-\dk+1},\cdots,S_{N}\in\mathcal{S}}
	    A_{N-\dk}(S_N,\cdots,S_{N-\dk+1})
	    \dyad{s_N^+}{s_N^-}\Bigg\Vert\\
	    \leqslant& \Bigg\Vert \sum_{D=0}^{\overline{D}}
		\sum_{(r^+, r^-)\in\mathcal{S}^2} 
		\sum_{\sgn\in\{+,-\}^D}
		\int_{\triangle_T^{(D)}}
		 A_{(r^+,r^-)}^{D,\sgn}
		 \left(t,[\tau_1,\cdots,\tau_D]\right) 
		 \dyad{r_f^+}{r_f^-}
		 \dif \boldsymbol{\tau} \\
		 &\quad -
		 \sum_{S_{N-\dk+1},\cdots,S_{N}\in\mathcal{S}}
	    A_{N-\dk}(S_N,\cdots,S_{N-\dk+1})
	    \dyad{s_N^+}{s_N^-}\Bigg\Vert + \varepsilon\\
	    \leqslant& 
	    \sum_{D=0}^{\overline{D}}
		\sum_{(r^+, r^-)\in\mathcal{S}^2} 
		\sum_{\sgn\in\{+,-\}^D}
		\Bigg\vert 
		\int_{\triangle_T^{(D)}}
		 A_{(r^+,r^-)}^{D,\sgn}
		 \left(t,[\tau_1,\cdots,\tau_D]\right) 
		 \dif \boldsymbol{\tau}\\
		 &\quad -
		 \sum_{\substack{S_{N-\dk+1},\cdots,S_{N-1}\in\mathcal{S} \\ S_k=(r_f^+,r_f^-), \text{number of spin flips}=D}}
	    A_{N-\dk}(S_N,\cdots,S_{N-\dk+1})
	    \Bigg\vert + 2\varepsilon \\
	    \leqslant& \sum_{D=0}^{\overline{D}} 2^{D+2}\frac{(2T)^D+1}{D!}\varepsilon + 2\varepsilon
	    \leqslant (4\e^{4T}+4\e^{2}+2)\varepsilon.
    \end{align*}
    The proof is completed as $\varepsilon$ is arbitrary.
    \end{proof}
\end{theorem}
With the reduced density matrix $\tilde{\rho}(t)$,
 any observable $O$ can be computed by
\begin{equation*}
    \expval{O(t)} = \tr(\rho_s(t)O).
\end{equation*}
In the following discussion,
 we mainly focus on the case with $O = \sigma_z = \mathrm{diag}(1,-1)$,
 and therefore
\begin{equation*}
    \expval{\sigma_z(t)} =
    \expval{\rho_s(t)}{+1}
    - \expval{\rho_s(t)}{-1}.
\end{equation*}


\section{Differential Equations for Path Integrals} 
\label{sec:quapi_pde}
We are now ready to formulate the equations for the functions $A_{(r^+,r^-)}^{D,\sgn} \left(t,[\tau_1,\cdots,\tau_D]\right)$ based on the i-QuAPI iteration \eqref{propagation}. While the resulting equations hold for $(\tau_1, \cdots, \tau_D) \in \triangle_T^{(D)}$ with $\triangle_T^{(D)}$ being an open set, they have to be completed by supplementing with appropriate boundary conditions. Moreover, as $D$ can take any positive integer, the form $A_{(r^+,r^-)}^{D,\sgn} \left(t,[\tau_1,\cdots,\tau_D]\right)$ actually includes infinite functions. In practice, one has to truncate the system and apply reasonable closure to make it numerically solvable. These topics will be discussed separately in the following subsections.

\subsection{Formulation of the Differential Equations}
With \Cref{thm_propagator},
 QuAPI partial differential equation
 can be derived in the domain $\triangle_T^{(D)}$
 for a spin-boson system.
The boundary case
 will be discussed in \Cref{boundaryCondition}
 in the next section.

In this section,
 a PDE based on (\ref{propagation}) for a two-state system is derived.
As is discussed above,
 $A$'s will be expressed as the form (\ref{notationA}). 

\begin{theorem}
    In the spin-boson model,
     the following partial differential equation holds
     for $A$ defined by \eqref{Asimplified}
     with the assumption that $A_{(r^+,r^-)}^{D,\sgn}(t,[\tau_1,\cdots,\tau_D])$ is differentiable with respect to $t$ and $\tau_1$.
	\begin{equation}
      \begin{split}
		\frac{\partial}{\partial t} 
		A_{(r^+,r^-)}^{D,\sgn}\left(t,[\tau_1,\cdots,\tau_D]\right)
		=& -W_{(r^+,r^-)}^{D,\sgn}\left([\tau_1,\cdots,\tau_D]\right)
		A_{(r^+,r^-)}^{D,\sgn}\left(t,[\tau_1,\cdots,\tau_D]\right) \\
		&+ A_{(r^+,\hat{r}^-)}^{D+1,[-,\sgn]}
		\left(t,[0,\tau_1,\cdots,\tau_D]\right)
		+ A_{(\hat{r}^+,r^-)}^{D+1,[+,\sgn]}
		\left(t,[0,\tau_1,\cdots,\tau_D]\right) \\
		&+ \frac{\partial}{\partial \tau_1} 
		A_{(r^+,r^-)}^{D,\sgn}\left(t,[\tau_1,\cdots,\tau_D]\right)
      \end{split} 
      \label{PDEoriginal}
	\end{equation}
	where $W_{(r^+,r^-)}^{D,\sgn}\left([\tau_1,\cdots,\tau_D]\right)$
	 is defined in (\ref{expressionW})
	 and $\hat{r}$ means the state different from $r$,
	 that is, $\widehat{ + 1} =  - 1  $
	 and $\widehat{ - 1} =  + 1  $.
	Note that when $D=0$,
	 the derivative of $A$ with respect to $\tau_1$ is zero.
	 \begin{proof}
	    We take the derivatives directly based on 
	     the expression \eqref{Asimplified}.
	    As the integral domain $\triangle_t^{(\tilde{D})}$ is dependent on $t$,
	     we first consider the derivative over the integral domain
	     for general function $f$:
	    \begin{equation}
	    \begin{split}
	        \frac{\partial}{\partial t} \int_{\triangle_t^{(\tilde{D})}}
	        f(t,\tilde{\tau}_1,\cdots,\tilde{\tau}_{\tilde{D}}) \dif\tilde{\boldsymbol{\tau}}
	        =& \int_{\triangle_t^{(\tilde{D})}}\frac{\partial}{\partial t}
	        f(t,\tilde{\tau}_1,\cdots,\tilde{\tau}_{\tilde{D}}) \dif \tilde{\boldsymbol{\tau}}\\
	        &+ \int_{\triangle_t^{(\tilde{D}-1)}} f(t,\tilde{\tau_1},\cdots,\tilde{\tau}_{\tilde{D}-1},
	        t-\tilde{\tau}_1-\cdots-\tilde{\tau}_{\tilde{D}-1})
	        \dif\tilde{\tau}_1\cdots\dif\tilde{\tau}_{\tilde{D}-1}.
	    \end{split}
	    \label{derivativeOfDomain}
	    \end{equation}
	    Especially, when $\tilde{D}=0$, the second term in \eqref{derivativeOfDomain} vanishes.
	    Therefore, 
	    \begin{equation}
	    \begin{split}
	        &\frac{\partial}{\partial t} A_{(r^+,r^-)}^{D,\sgn}(t,[\tau_1,\cdots,\tau_D]) \\
	        =& \sum_{\tilde{D}=1}^\infty
            \sum_{(\tilde{r}_0^+,\tilde{r}_0^-)\in\mathcal{S}^2}
            \sum_{ \widetilde{\sgn} \in \{+,-\}^{\tilde{D}} }
            \int_{\triangle_t^{(\tilde{D}-1)}}\dif \tilde{\tau}_1
            \cdots\tilde{\tau}_{\tilde{D}-1}
            \left(X(g)
            \delta_{\tilde{r}_{\tilde{D}}^+}^{r_0^+}
            \delta_{\tilde{r}_{\tilde{D}}^-}^{r_0^-}
            Y(h)
            \e^{Z(g,h)}\right)\Big\vert_{\tilde{\tau}_{\tilde{D}}=t-\tilde{\tau}_1-\cdots-\tilde{\tau}_{\tilde{D}-1}} \\
            &+ \sum_{\tilde{D}=0}^\infty
            \sum_{(\tilde{r}_0^+,\tilde{r}_0^-)\in\mathcal{S}^2}
            \sum_{ \widetilde{\sgn} \in \{+,-\}^{\tilde{D}} }
            \int_{\triangle_t^{(\tilde{D})}}\dif \tilde{\boldsymbol{\tau}}
            \frac{\partial}{\partial t}X(g)
            \delta_{\tilde{r}_{\tilde{D}}^+}^{r_0^+}
            \delta_{\tilde{r}_{\tilde{D}}^-}^{r_0^-}
            Y(h)
            \e^{Z(g,h)} \\
            &+\sum_{\tilde{D}=0}^\infty
            \sum_{(\tilde{r}_0^+,\tilde{r}_0^-)\in\mathcal{S}^2}
            \sum_{ \widetilde{\sgn} \in \{+,-\}^{\tilde{D}} }
            \int_{\triangle_t^{(\tilde{D})}}\dif \tilde{\boldsymbol{\tau}}
            X(g)
            \delta_{\tilde{r}_{\tilde{D}}^+}^{r_0^+}
            \delta_{\tilde{r}_{\tilde{D}}^-}^{r_0^-}
            Y(h)
            \frac{\partial}{\partial t}Z(g,h)\e^{Z(g,h)}
	    \end{split}
	    \label{partialApartialt}
	    \end{equation}
	    In addition, we take the partial derivative with respect to $\tau_1$.
	    \begin{equation*}
        \begin{split}
            &\frac{\partial}{\partial \tau_1}
            A_{(r^+,r^-)}^{D,\sgn}(t,[\tau_1,\cdots,\tau_D]) \\
            =& \sum_{\tilde{D}=0}^\infty \sum_{(\tilde{r}_0^+,\tilde{r}_0^-)\in\mathcal{S}^2}
            \sum_{\widetilde{\sgn}\in\{+,-\}^{\tilde{D}}}
            \int_{\triangle_t^{(\tilde{D})}}\dif \tilde{\boldsymbol{\tau}}
            X(g)
            \delta_{\tilde{r}_{\tilde{D}}^+}^{r_0^+}
            \delta_{\tilde{r}_{\tilde{D}}^-}^{r_0^-}
            \frac{\partial}{\partial \tau_1} Y(h)
            \e^{Z(g,h)} \\
            &+ \sum_{\tilde{D}=0}^\infty \sum_{(\tilde{r}_0^+,\tilde{r}_0^-)\in\mathcal{S}^2}
            \sum_{\widetilde{\sgn}\in\{+,-\}^{\tilde{D}}}
            \int_{\triangle_t^{(\tilde{D})}}\dif \tilde{\boldsymbol{\tau}}
            X(g)
            \delta_{\tilde{r}_{\tilde{D}}^+}^{r_0^+}
            \delta_{\tilde{r}_{\tilde{D}}^-}^{r_0^-}
            Y(h)
            \frac{\partial}{\partial \tau_1} Z(g,h) \e^{Z(g,h)}.
        \end{split}
        \end{equation*}
        For any given path $g$,
        \begin{equation*}
        \begin{split}
            &\frac{\partial}{\partial t}X(g)
            \delta_{\tilde{r}_{\tilde{D}}^+}^{r_0^+}
            \delta_{\tilde{r}_{\tilde{D}}^-}^{r_0^-}
            Y(h)
            \e^{Z(g,h)}
            - X(g)
            \delta_{\tilde{r}_{\tilde{D}}^+}^{r_0^+}
            \delta_{\tilde{r}_{\tilde{D}}^-}^{r_0^-}
            \frac{\partial}{\partial \tau_1} Y(h)
            \e^{Z(g,h)} \\
            = & -\ii \left(\mel{\tilde{r}_{\tilde{D}}^+}{H_0}{\tilde{r}_{\tilde{D}}^+} -\mel{\tilde{r}_{\tilde{D}}^-}{H_0}{\tilde{r}_{\tilde{D}}^-}\right) X(g)
            \delta_{\tilde{r}_{\tilde{D}}^+}^{r_0^+}
            \delta_{\tilde{r}_{\tilde{D}}^-}^{r_0^-}
            Y(h) \e^{Z(g,h)} \\
            &- \left( -\ii \big(
            \mel{r_0^+}{H_0}{r_0^+} - \mel{r_0^-}{H_0}{r_0^-}
            \big)
            + \ii
            \big(\mel{r_D^+}{H_0}{r_D^-} - \mel{r_D^-}{H_0}{r_D^-}\big)
            \right)
            X(g) 
            \delta_{\tilde{r}_{\tilde{D}}^+}^{r_0^+}
            \delta_{\tilde{r}_{\tilde{D}}^-}^{r_0^-}
            Y(h)
            \e^{Z(g,h)} \\
            = & -\ii \big(\mel{r_D^+}{H_0}{r_D^-} - \mel{r_D^-}{H_0}{r_D^-}\big)
            X(g) 
            \delta_{\tilde{r}_{\tilde{D}}^+}^{r_0^+}
            \delta_{\tilde{r}_{\tilde{D}}^-}^{r_0^-}
            Y(h)
            \e^{Z(g,h)}.
        \end{split}
        \end{equation*}
        If $\delta_{\tilde{r}_{\tilde{D}}^+}^{r_0^+}
            \delta_{\tilde{r}_{\tilde{D}}^-}^{r_0^-} = 0$,
        both $X(g)
            \delta_{\tilde{r}_{\tilde{D}}^+}^{r_0^+}
            \delta_{\tilde{r}_{\tilde{D}}^-}^{r_0^-}
            Y(h)
            \frac{\partial}{\partial t}Z(g,h)\e^{Z(g,h)}$
        and $X(g)
            \delta_{\tilde{r}_{\tilde{D}}^+}^{r_0^+}
            \delta_{\tilde{r}_{\tilde{D}}^-}^{r_0^-}
            Y(h)
            \frac{\partial}{\partial \tau_1} Z(g,h) \e^{Z(g,h)}$
        are zero.
        If $\delta_{\tilde{r}_{\tilde{D}}^+}^{r_0^+}
            \delta_{\tilde{r}_{\tilde{D}}^-}^{r_0^-} = 1$,
        we have $(\tilde{r}_{\tilde{D}}^+,\tilde{r}_{\tilde{D}}^-) = (r_0^+,r_0^-)$,
        we have
        \begin{equation*}
        \begin{split}
            &\frac{\partial}{\partial t}Z(g,h)
            - \frac{\partial}{\partial \tau_1}Z(g,h) \\
            =& \lim_{\dt\rightarrow 0} \frac{1}{\dt}\Bigg(
            -\int_{t+T}^{t+T+\dt}\dif x_1 \int_{x_1-T}^{t+\dt} \dif x_2
            (h^+(x_1-t-\dt)-h^-(x_1-t-\dt)) \\
            &\qquad\qquad(\tilde{r}_{\tilde{D}}^+ \tilde{\eta}(x_1-x_2)
            - \tilde{r}_{\tilde{D}}^- \tilde{\eta}^*(x_1-x_2)) \\
            &\qquad - 
            \int_{t+T}^{t+T+\dt}\dif x_1 \int_{t+\dt}^{x_1} \dif x_2
            (h^+(x_1-t-\dt)-h^-(x_1-t-\dt)) \\
            &\qquad\qquad(h^+(x_2-t-\dt) \tilde{\eta}(x_1-x_2)
            - h^-(x_2-t-\dt) \tilde{\eta}^*(x_1-x_2))
            \Bigg) \\
            =& -\int_{0}^{T}\dif x_2
            \left(h^+(T) - h^-(T))(h^+(x_2)\tilde{\eta}(T-x_2)
            -h^-(x_2)\tilde{\eta}^*(T-x_2)\right).
        \end{split}
        \end{equation*}
        For the first term in \eqref{partialApartialt}, consider the following two quantities:
        \begin{gather}
            \sum_{\tilde{D}=1}^\infty
            \sum_{(\tilde{r}_0^+,\tilde{r}_0^-)\in\mathcal{S}^2}
            \sum_{ \widetilde{\sgn} \in \{+,-\}^{\tilde{D}} }
            \int_{\triangle_t^{(\tilde{D}-1)}}\dif \tilde{\tau}_1
            \cdots\tilde{\tau}_{\tilde{D}-1}
            \left(X(g)
            \delta_{\tilde{r}_{\tilde{D}}^+}^{r_0^+}
            \delta_{\tilde{r}_{\tilde{D}}^-}^{r_0^-}
            Y(h)
            \e^{Z(g,h)}\right)\Big\vert_{\tilde{\tau}_{\tilde{D}}=t-\tilde{\tau}_1-\cdots-\tilde{\tau}_{\tilde{D}-1}}; 
            \label{restTerm}\\
            A_{(r^+,\hat{r}^-)}^{D+1,[-,\sgn]}
    		\left(t,[0,\tau_1,\cdots,\tau_D]\right)
    		+ A_{(\hat{r}^+,r^-)}^{D+1,[+,\sgn]}
    		\left(t,[0,\tau_1,\cdots,\tau_D]\right).
    		\label{resultRestTerm}
        \end{gather}
        The first term in \eqref{partialApartialt}
         involves letting 
         $\tilde{\tau}_{\tilde{D}} = t-\tilde{\tau}_1-\cdots-\tilde{\tau}_{\tilde{D}-1}$,
         which means $(\tilde{\tau}_1,\cdots,\tilde{\tau}_{\tilde{D}})\in\partial\triangle_t^{(D)}$ and therefore,
         we need to regard it as a limit case.
        For any $\tilde{D}\geqslant 1$, $(\tilde{r}_0^+,\tilde{r}_0^-)\in\mathcal{S}^2$,
        $\widetilde{\sgn}\in\{+,-\}^{\tilde{D}}$ with $\tilde{\sgn}_{\tilde{D}}$ being $+$, define the path $g:[0,t)\rightarrow\mathcal{S}^2$ by \eqref{function_g}.
        In addition, define $g':[0,t)\rightarrow\mathcal{S}^2$ by
        \begin{equation*}
            g'(\tau) = \begin{cases}
            g(\tau), &\text{ if } \tau\in[0,\tilde{\tau}_1+\cdots+\tilde{\tau}_{\tilde{D}}) \\
            \lim_{s\rightarrow(\tilde{\tau}_1+\cdots,\tilde{\tau}_{\tilde{D}})^-} g(\tau), &\text{ if } \tau\in[\tilde{\tau}_1+\cdots+\tilde{\tau}_{\tilde{D}},t)
            \end{cases}.
        \end{equation*}
        Compare with path $g$, path $g'$ has one fewer spin flip on the positive branch at time $\tilde{\tau}_1+\cdots+\tilde{\tau}_{\tilde{D}}$.
        In addition, defined a path $h_{\dt}:[0,T)\rightarrow\mathcal{S}^2$ by
        \begin{equation*}
            h_{\dt}(\tau) = \begin{cases}
                (\hat{r}_0^+,r_0^-), & \text{ if } \tau\in[0,\dt)\\
                h(\tau), &\text{ if }
                \tau\in[\dt,T)
            \end{cases}.
        \end{equation*}
        If $\delta_{\tilde{r}_{\tilde{D}}^+}^{r_0^+}
            \delta_{\tilde{r}_{\tilde{D}}^-}^{r_0^-}=0$,
         then we naturally have
        \begin{equation}
            \lim_{\tilde{\tau}_{\tilde{D}}\rightarrow 
            t-\tilde{\tau}_1-\cdots-\tilde{\tau}_{\tilde{D}-1}}
            X(g)
            \delta_{\tilde{r}_{\tilde{D}}^+}^{r_0^+}
            \delta_{\tilde{r}_{\tilde{D}}^-}^{r_0^-}
            Y(h)\e^{Z(g,h)}
            = \lim_{\dt\rightarrow 0} X(g')
            \delta_{\tilde{r}_{\tilde{D}-1}^+}^{\hat{r}_0^+}
            \delta_{\tilde{r}_{\tilde{D}-1}^-}^{r_0^-}
            Y(h_{\dt})\e^{Z(g',h_{\dt})}
            \label{XYeZ}
        \end{equation}
        as they are both zero.
        If $\delta_{\tilde{r}_{\tilde{D}}^+}^{r_0^+}
            \delta_{\tilde{r}_{\tilde{D}}^-}^{r_0^-} = 1$,
        both $\delta_{\tilde{r}_{\tilde{D}}^+}^{r_0^+}
            \delta_{\tilde{r}_{\tilde{D}}^-}^{r_0^-}=
            \delta_{\tilde{r}_{\tilde{D}-1}^+}^{\hat{r}_0^+}
            \delta_{\tilde{r}_{\tilde{D}-1}^-}^{r_0^-} = 1$
        and we just omit the $\delta$ symbols in the following statements.
        We consider the quotient of $X(g)Y(h)\e^{Z(g,h)}$ 
         and $X(g')Y(h_{\dt})\e^{Z(g,h)}$.
        By definition,
        \begin{equation*}
            \frac{X(g)}{X(g')}
            = \frac{\e^{-\ii\left(
            \mel{\tilde{r}_{\tilde{D}}^+}{H_0}{\tilde{r}_{\tilde{D}}^+} - 
            \mel{\tilde{r}_{\tilde{D}}^-}{H_0}{\tilde{r}_{\tilde{D}}^-}
            \right)\left(t-\sum_{j=1}^{\tilde{D}}\tilde{\tau}_j\right)}
            \left(-\ii\mel{\tilde{r}_{\tilde{D}}^+}{H_0}{\tilde{r}_{\tilde{D}-1}^+}\right)
            }
            {
            \e^{-\ii\left(
            \mel{\tilde{r}_{\tilde{D}-1}^+}{H_0}{\tilde{r}_{\tilde{D}-1}^+} - 
            \mel{\tilde{r}_{\tilde{D}-1}^-}{H_0}{\tilde{r}_{\tilde{D}-1}^-}\left(t-\sum_{j=1}^{\tilde{D}}\tilde{\tau}_j\right)
            \right)}
            }
        \end{equation*}
        and therefore
        \begin{equation*}
            \lim_{\tilde{\tau}_D\rightarrow t-\tilde{\tau}_1-\cdots-\tilde{\tau}_{\tilde{D}-1}}
            \frac{X(g)}{X(g')}
            = -\ii\mel{\tilde{r}_{\tilde{D}}^+}{H_0}{\tilde{r}_{\tilde{D}-1}^+}.
        \end{equation*}
        Similarly
        \begin{equation*}
            \frac{Y(h)}{Y(h_{\dt})}
            = \frac{\e^{-\ii\left(\mel{r_0^+}{H_0}{r_0^+} - \mel{r_0^-}{H_0}{r_0^-}\right)\dt}}
            {\e^{\left(\mel{\hat{r}_0^+}{H_0}{\hat{r}_0^+} - \mel{r_0^-}{H_0}{r_0^-}\right)\dt}
            \left(-\ii\mel{\hat{r}_0^+}{H_0}{r_0^+}\right)}
        \end{equation*}
        and therefore,
        \begin{equation*}
            \lim_{\dt\rightarrow 0}\frac{Y(h)}{Y(h_{\dt})} = \frac{1}{-\ii\mel{\hat{r}_0^+}{H_0}{r_0^+}}.
        \end{equation*}
        As for the term $Z$, it is clear that
        \begin{equation*}
            \lim_{\substack{\tilde{\tau}_D \rightarrow t- \tilde{\tau}_1 -\cdots- \tilde{\tau}_{\tilde{D}-1} \\ \dt \rightarrow 0}}\frac{Z(g,h)}{Z(g',h_{\dt})} = 1.
        \end{equation*}
        Therefore, \eqref{XYeZ} also holds for the case $\delta_{\tilde{r}_{\tilde{D}}^+}^{r_0^+}
            \delta_{\tilde{r}_{\tilde{D}}^-}^{r_0^-}=1$.
        The proof is similar for the case when $\widetilde{\sgn}_{\tilde{D}}$ is `$-$'.
        Take the sum over $\tilde{D}$, $(\tilde{r}_0^+,\tilde{r}_0^-)$ and $\widetilde{\sgn}$, we conclude that \eqref{restTerm} equals to \eqref{resultRestTerm}.
        Combine all the results and \eqref{PDEoriginal} holds
         with $W$ defined in \Cref{thm_propagator}.
        Note that it is easy to check when $D = 0$,
         \eqref{PDEoriginal} holds by regarding the derivative of $A$ with respect to $\tau_1$ being 0. 
	 \end{proof}
\end{theorem}

Although only spin-boson model is considered
 in the derivation of the PDE,
 for more complicated cases,
 the derivation keeps the same
 but more terms should be taken into consideration.

\subsection{Boundary Conditions}
\label{sec:boundaryCondition}
The equation \eqref{PDEoriginal} turns out 
 to be a hyperbolic equation 
 with linear advection in the $\tau_1$ direction, 
 and therefore requires initial and boundary conditions 
 to complete the problem statement. 
While the initial condition has been provided in \eqref{Asimplified} by choosing $t=0$, 
 the boundary conditions are still to be formulated. 
The advection in \eqref{PDEoriginal} shows 
 that the boundary condition should be provided at largest $\tau_1$ for all possible choices of other arguments. 
Due to the constraint \eqref{DefnSimplex}, 
 the boundary condition should specify the function values at $\tau_1 = T - \tau_2 - \cdots - \tau_D$. Note that in this case, the corresponding path is not defined in the formula \eqref{contPath} since $(\tau_1,\cdots,\tau_D)$ locates on the boundary of $\triangle_T^{(D)}$. In fact, this corresponds to the path segment $h$ with
 one or more spin flips
 occurring exactly at time $T$. 
 In this special situation, the value of $A_{(r^+,r^-)}^{D,\sgn}(t,
 \boldsymbol{\tau})$ can be expressed by the value of $A(t,h')$ with only $D-1$ spin flips in $h'$. The details are given in the following theorem:
 
\begin{theorem}
\label{boundaryCondition}
For $(\tau_1,\cdots,\tau_{D-1})\in\triangle_{T}^{(D-1)}$, 
 $\sgn\in\{+,-\}^D$ and $(r^+,r^-)\in\mathcal{S}^2$,
 the following equality holds: 
\begin{equation}
\begin{split}
    &A_{(r^+,r^-)}^{D,\sgn}(t,[\tau_1,\cdots,\tau_{D-1},T-\tau_1-\cdots-\tau_{D-1}]) \\
    ={} & A_{(r^+,r^-)}^{D-1,\sgn_{1:D-1}}(t,[\tau_1,\cdots,\tau_{D-1}])
    \left(-\ii\mel{r_f^+}{H_0}{\hat{r}_f^+}
    \delta_{\sgn_D}^+
    + \ii \mel{\hat{r}_f^-}{H_0}{r_f^-}
    \delta_{\sgn_D}^-\right).
\end{split}
\label{eq_boundaryCondition}
\end{equation}
where $r_f^{\pm} = r_f^{\pm}(r^+,r^-,D,\sgn)$
 is defined by \eqref{finalState}.
In this equation, $\sgn_{k_1:k_2}$ is the $k_1$-th to $k_2$-th components of $\sgn$.
\begin{proof}
    It is clear that $(\tau_1,\cdots,\tau_{D-1},T-\tau_1-\cdots-\tau_{D-1})\in\partial\triangle_T^{(D)}$.
    According to \Cref{boundaryADefn},
     we have the following limit:
    \begin{equation*}
        A_{(r^+,r^-)}^{D,\sgn}(t,[\tau_1,\cdots,\tau_{D-1},T-\tau_1-\cdots-\tau_{D-1}])
        = \lim_{\tau_D\rightarrow (T-\tau_1-\cdots-\tau_{D-1})^-}
        A_{(r^+,r^-)}^{D,\sgn}(t,[\tau_1,\cdots,\tau_D]).
    \end{equation*}
    For simplicity, given $(\tau_1, \cdots, \tau_D) \in \triangle_T^{(D)}$, below we use $h$ to denote the path segment defined by \eqref{contPath} and use $h'$ to denote the path segment defined by removing the last spin flip in $h$:
    \begin{displaymath}
    h'(\tau) = \begin{cases}
      h(\tau), & \text{if } \tau \in [0,\tau_1+\cdots+\tau_D), \\
      \lim\limits_{s \rightarrow (\tau_1 + \cdots + \tau_D)^-} h(s), & \text{if } \tau \in [\tau_1 + \cdots + \tau_D, T).
    \end{cases}
    \end{displaymath}
    Then we have
    $A_{(r^+,r^-)}^{D,\sgn}(t,[\tau_1,\cdots,\tau_D]) = A(t,h)$
    and
    $A_{(r^+,r^-)}^{D-1,\sgn_{1:D-1}}(t,[\tau_1,\cdots,\tau_{D-1}]) = A(t,h')$. Since
    the only difference between this two paths is the last segment with length $T-\tau_1-\cdots-\tau_D$,
    which vanishes as $\tau_D\rightarrow(T-\tau_1-\cdots-\tau_{D-1})^-$,
    we conclude that
    \begin{equation*}
        \lim_{\tau_D\rightarrow(T-\tau_1-\cdots-\tau_{D-1})^-}
        Z(g,h) = Z(g,h')
    \end{equation*}
    for any path $g:[0,t)\rightarrow \mathcal{S}^2$.
    In addition,
    \begin{equation*}
    \begin{split}
        &\lim_{\tau_D\rightarrow(T-\tau_1-\cdots-\tau_{D-1})^-} \frac{Y(h)}{Y(h')} \\
        ={}& \lim_{\tau_D\rightarrow(T-\tau_1-\cdots-\tau_{D-1})^-}\frac{
        \e^{-\ii\left(\mel{r_D^+}{H_0}{r_D^+}
        -\mel{r_D^-}{H_0}{r_D^-}\right)
        \left(T-\sum_{j=1}^D\tau_j\right)}
        \left(-\ii\mel{r_D^+}{H_0}{r_{D-1}^+}\delta_{\sgn_D}^+ 
        + \ii \mel{r_{D-1}^-}{H_0}{r_D^-}\delta_{\sgn_D}^-\right)}
        {\e^{-\ii\left(\mel{r_{D-1}^+}{H_0}{r_{D-1}^+}
        -\mel{r_{D-1}^-}{H_0}{r_{D-1}^-}\right)
        \left(T-\sum_{j=1}^D\tau_j\right)}} \\
        ={} & -\ii\mel{r_D^+}{H_0}{r_{D-1}^+}\delta_{\sgn_D}^+ 
        + \ii \mel{r_{D-1}^-}{H_0}{r_D^-}
        \delta_{\sgn_D}^-.
    \end{split}
    \end{equation*}
    Therefore, for any path $g$,
    \begin{equation*}
        \lim_{\tau_D\rightarrow(T-\tau_1-\cdots-\tau_{D-1})^-} \frac{X(g)\delta_{\tilde{r}_{\tilde{D}}^+}^{r_0^+}
        \delta_{\tilde{r}_{\tilde{D}}^-}^{r_0^-}Y(h)\e^{Z(g,h)}}{X(g)\delta_{\tilde{r}_{\tilde{D}}^+}^{r_0^+}
        \delta_{\tilde{r}_{\tilde{D}}^-}^{r_0^-}Y(h')\e^{Z(g,h')}}
        = \left(-\ii\mel{r_D^+}{H_0}{r_{D-1}^+}\delta_{\sgn_D}^+ 
        + \ii \mel{r_{D-1}^-}{H_0}{r_D^-}
        \delta_{\sgn_D}^-\right)
    \end{equation*}
    and consequently, the conclusion holds
    as $r_f^\pm = r_D^\pm$.
\end{proof}
\end{theorem}
In conclusion,
 the boundary conditions for the functions with $D$ spin flips
 are determined by the case with one less spin flip.
Note that when we allow $(\tau_1, \cdots, \tau_D)$ to take any values on $\partial \triangle_T^{(D)}$, there may be more than one spin flips at time $T$,
which happens if $\tau_1 + \cdots + \tau_D = T$ and $\tau_D = 0$.
 In this case, the boundary condition can be determined by
 applying \eqref{eq_boundaryCondition} repeatedly. In general, for $k \leqslant D$, if $\tau_1 + \cdots + \tau_k = T$, we have
 \begin{displaymath}
 \begin{split}
 &A_{(r^+,r^-)}^{D,\sgn}(t,[\tau_1, \cdots, \tau_k, 0, \cdots, 0]) \\
 = &A_{(r^+,r^-)}^{k-1,\sgn_{1:k-1}}(t,[\tau_1,\cdots,\tau_{k-1}]) \prod_{j=k}^D 
 \left(-\ii\mel{r_j^+}{H_0}{\hat{r}_j^+}
 \delta_{\sgn_j}^+
 + \ii \mel{\hat{r}_j^-}{H_0}{r_j^-}
 \delta_{\sgn_j}^-\right),
\end{split}
 \end{displaymath}
 where $r_j$ is the $j$-th state in the path segment:
 \begin{displaymath}
   r_j = r_f(r^+, r^-, j, \sgn_{1:j}).
 \end{displaymath}

By now, with the equation \eqref{PDEoriginal} and the boundary conditions \eqref{eq_boundaryCondition}, a complete system of infinite partial differential equations has been formulated. 

\subsection{Truncation and Closure of the System}
\label{ClosureOfTheSystem}
The system \eqref{PDEoriginal} includes infinite equations since $D$ can take the value of any non-negative integer. However, for practical numerical computation, we can only solve finite equations, which requires a truncation of the system. Physically, due to the finite spin flipping rate, the probability that the system undergoes $D$ spin flips generally follows the Poisson distribution. Therefore when $D$ is large, the contribution from $A_{(r^+,r^-)}^{D, \sgn}$ can be regarded as small, so that it is reasonable to discard evolution equations for $A_{(r^+,r^-)}^{D, \sgn}$ with large $D$. This can also be justified mathematically since
\begin{equation*}
    \left\Vert\sum_{D = \overline{D}+1}^{\infty} \sum_{(r^+,r^-)\in\mathcal{S}^2}
    \sum_{\sgn\in\{+,-\}^{D}}
    \int_{\triangle_T^{(D)}}
    A_{(r^+,r^-)}^{D,\sgn}(t,[\tau_1,\cdots,\tau_D])
    \dyad{r_f^+}{r_f^-}
    \dif{\boldsymbol{\tau}}\right\Vert
    \rightarrow 0, \quad \text{as}\quad \overline{D}\rightarrow\infty,
\end{equation*}
which has been shown in the proof of \Cref{densityMatrix}. Below in our numerical method, we will only focus on the equations of $A_{(r^+,r^-)}^{D,\sgn}$ with $D \in \{0,1,\cdots,D_{\max}\}$ for some $D_{\max}$ chosen suitable for the problem.

However, the equation \eqref{PDEoriginal} shows that the evolution of $A_{(r^+,r^-)}^{D_{\max}, \sgn}$ depends on the values of $A_{(r^+,r^-)}^{D_{\max}+1, \sgn}$, meaning that our truncation does not form a closed system, which requires us to impose an appropriate closure. Precisely speaking, we need to provide some predictions of $A_{(r^+,r^-)}^{D_{\max}+1, \sgn}$ based on the values of $A_{(r^+,r^-)}^{D, \sgn}$ with $D \leqslant D_{\max}$. 
Our approach is to estimate the values of $A_{(r^+,r^-)}^{D_{\max}+1,\sgn}$ via interpolations
based on \Cref{boundaryCondition} and the following proposition.
\begin{proposition}
\label{addtwospinflips}
For any $(r^+,r^-), D \geqslant 2$ and $\sgn$
 with $\sgn_k = \sgn_{k+1}$, if $\tau_k=0$,
\begin{equation*}
    \begin{split}
    &A_{(r^+,r^-)}^{D,\sgn}
    (t,[\tau_1,\cdots,\tau_D])\\
    =& - A_{(r^+,r^-)}^{D-2,[\sgn_{1:k-1}, \sgn_{k+2:D}]}(t,[\tau_1,\cdots,\tau_{k-2},\tau_{k-1}+\tau_{k+1}, \tau_{k+2},\cdots,\tau_D])
    \mel{+1}{H_0}{-1}\mel{-1}{H_0}{+1}.
    \end{split}
\end{equation*}
\begin{proof}
    It is clear that $(\tau_1,\cdots,\tau_D)\in \partial\triangle_T^{(D)}$.
    By \Cref{boundaryADefn},
     the following limit holds:
    \begin{equation*}
    \begin{split}
        &A_{(r^+,r^-)}^{D,\sgn}(t,[\tau_1,\cdots,\tau_D])
        = \lim_{\dt\rightarrow 0^-}
        A_{(r^+,r^-)}^{D,\sgn}(t,[\tau_1,\cdots,\tau_{k-1},\dt,\tau_{k+1}-\dt,\tau_{k+2},\cdots,\tau_D])
        := \lim_{\dt\rightarrow 0^-} A(t,h_{\dt}),
    \end{split}
    \end{equation*}
    and for simplicity,
     we denote the path of
    $ A_{(r^+,r^-)}^{D-2,[\sgn_{1:k-1}, \sgn_{k+2:D}]}(t,[\tau_1,\cdots,\tau_{k-2},\tau_{k-1}+\tau_{k+1}, \tau_{k+2},\cdots,\tau_D])$
     as $h'$.
    The only difference between the path $h_{\dt}$ and $h$
     is in the interval $\left[\sum_{j=1}^k,\tau_j,\sum_{j=1}^k \tau_j+\dt\right)$.
    Therefore, similar to the proof of \Cref{boundaryCondition}, we have
    $\lim_{\dt\rightarrow 0^-}Z(g,h_{\dt}) = Z(g,h')$
    for any path $g:[0,t)\rightarrow \mathcal{S}^2$.
    In addition, similar to the proof of \Cref{boundaryCondition}, we have
    \begin{equation*}
        \lim_{\dt\rightarrow 0}
        \frac{Y(h_{\dt})}{Y(h')}
        =-\mel{+1}{H_0}{-1}\mel{-1}{H_0}{+1}
    \end{equation*}
    Consequently, we have
    \begin{equation*}
        \lim_{\dt\rightarrow 0}\frac{X(g)\delta_{\tilde{r}_{\tilde{D}}^+}^{r_0^+}
        \delta_{\tilde{r}_{\tilde{D}}^-}^{r_0^-}Y(h_{\dt})Z(g,h_{\dt})}{X(g)\delta_{\tilde{r}_{\tilde{D}}^+}^{r_0^+}
        \delta_{\tilde{r}_{\tilde{D}}^-}^{r_0^-}Y(h')Z(g,h')} = -\mel{+1}{H_0}{-1}\mel{-1}{H_0}{+1}
    \end{equation*}
    for all path $g$ and the result holds.
\end{proof}
\end{proposition}

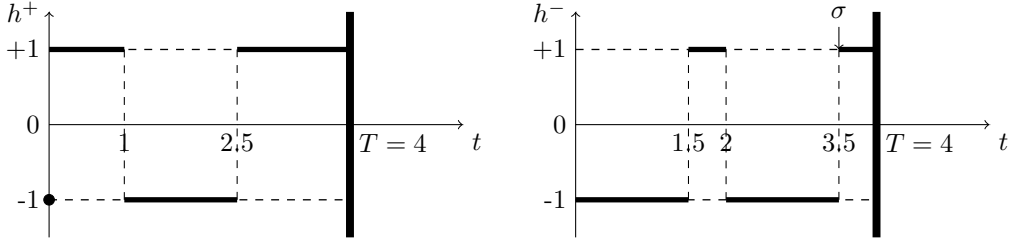
\begin{figure}[!ht]
    \centering
\begin{tikzpicture}
  \draw[->] (0,0) -- (5.5,0);
  \draw[->] (0,-1.5) -- (0,1.5);
  \draw[line width = 3pt] (4,-1.5) -- (4,1.5);
  \filldraw[black] (0,-1) circle (2pt);
  \draw[line width = 2pt] (0,1) -- (1,1);
  \draw[dashed] (1,1) -- (1,-1);
  \draw[line width = 2pt] (1,-1) -- (2.5,-1);
  \draw[dashed] (2.5,-1) -- (2.5,1);
  \draw[line width = 2pt] (2.5,1) -- (4,1);
  \draw[dashed, line width=0.5pt] (0,1) -- (4,1);
  \draw[dashed, line width=0.5pt] (0,-1) -- (4,-1);
  
  \filldraw[black] (0,0) circle node[anchor=east] {0};
  \filldraw[black] (1,0) circle node[anchor=north] {1};
  \filldraw[black] (2.5,0) circle node[anchor=north] {2.5};
  \filldraw[black] (4,0) circle node[anchor=north west] {$T=4$};
  \filldraw[black] (5.5,0) circle node[anchor=north west] {$t$};
  \filldraw[black] (0,1) circle node[anchor=east] {+1};
  \filldraw[black] (0,-1) circle node[anchor=east] {-1};
  \filldraw[black] (0,1.5) circle node[anchor=east] {$h^+$};
  
  \draw[->] (7,0) -- (12.5,0);
  \draw[->] (7,-1.5) -- (7,1.5);
  \draw[line width = 3pt] (11,-1.5) -- (11,1.5);
  \draw[line width = 2pt] (7,-1) -- (8.5,-1);
  \draw[dashed] (8.5,-1) -- (8.5,1);
  \draw[line width = 2pt] (8.5,1) -- (9,1);
  \draw[dashed] (9,1) -- (9,-1);
  \draw[line width = 2pt] (9,-1) -- (10.5,-1);
  \draw[dashed] (10.5,-1) -- (10.5,1);
  \draw[line width = 2pt] (10.5,1) -- (11,1);
  \draw[dashed, line width=0.5pt] (7,1) -- (11,1);
  \draw[dashed, line width=0.5pt] (7,-1) -- (11,-1);
  
  \filldraw[black] (7,0) circle node[anchor=east] {0};
  \filldraw[black] (8.5,0) circle node[anchor=north] {1.5};
  \filldraw[black] (9,0) circle node[anchor=north] {2};
  \filldraw[black] (10.5,0) circle node[anchor=north] {3.5};
  \filldraw[black] (11,0) circle node[anchor=north west] {$T=4$};
  \filldraw[black] (12.5,0) circle node[anchor=north west] {$t$};
  \filldraw[black] (7,-1) circle node[anchor=east] {-1};
  \filldraw[black] (7,1) circle node[anchor=east] {+1};
  \filldraw[black] (7,1.5) circle node[anchor=east] {$h^-$};
  \node (tl) at (10.5,1.5) {$\sigma$};
  \draw[->] (tl) -- (10.5,1);
\end{tikzpicture}
    \caption{The illustration of the path segment in $A_{(-1,-1)}^{6,[+,+,-,-,+,-]}(t,[0,1.0,0.5,0.5,0.5,1.0])$.}
    \label{fig:needToCompute}
\end{figure}

The proposition above indicates that if two spin flips occur at the same time on the same branch, the function value of this path segment can be represented using the function value of the path segment with both spin flips removed. Therefore, this proposition establishes the relationship between the function values for path segments with different values of $D$.
To demonstrate how this is applied to derive the closure of the system,
 we start with an example with $D_{\max} = 5$, and we try to estimate 
$A_{(-1,-1)}^{6,[+,+,-,-,+,-]}(t,[0,1.0,0.5,0.5,0.5,1.0])$
 (see \Cref{fig:needToCompute}).
To this end, we consider the time point of the last spin flip at $t = 3.5$ on the negative branch as a variable $\sigma$ while fixing all other spin flips.
 According to the structure of $h^-$, the value of $\sigma$ can take any number between $t=2$ and $t=4$.
When $\sigma = 2$, we obtain a path with two spin flips at the same time point, which is illustrated in \Cref{fig:STEP3FIG2}, and the corresponding value of $A(t,h)$ is represented by
 $A_{(-1,-1)}^{6,[+,+,-,-,-,+]}(t,[0,1.0,0.5,0.5,0,0.5])$.
Similarly, when $\sigma=4$,
 we get $A_{(-1,-1)}^{6,[+,+,-,-,+,-]}(t,[0,1.0,0.5,0.5,0.5,1.5])$ (see \Cref{fig:STEP3FIG1}). The idea of our closure is to use the following linear interpolation as an approximation:
\begin{equation*}
\begin{split}
     &A_{(-1,-1)}^{6,[+,+,-,-,+,-]}(t,[0,1.0,0.5,0.5,0.5,1.0]) \\
   \approx{} &\frac{1}{4} A_{(-1,-1)}^{6,[+,+,-,-,-,+]}(t,[0,1.0,0.5,0.5,0,0.5])
   +\frac{3}{4} A_{(-1,-1)}^{6,[+,+,-,-,+,-]}(t,[0,1.0,0.5,0.5,0.5,1.5]).
\end{split}
\end{equation*}
\begin{figure}[!ht]
    \centering
\begin{tikzpicture}
  \draw[->] (0,0) -- (5.5,0);
  \draw[->] (0,-1.5) -- (0,1.5);
  \draw[line width = 3pt] (4,-1.5) -- (4,1.5);
  \filldraw[black] (0,-1) circle (2pt);
  \draw[line width = 2pt] (0,1) -- (1,1);
  \draw[dashed] (1,1) -- (1,-1);
  \draw[line width = 2pt] (1,-1) -- (2.5,-1);
  \draw[dashed] (2.5,-1) -- (2.5,1);
  \draw[line width = 2pt] (2.5,1) -- (4,1);
  \draw[dashed, line width=0.5pt] (0,1) -- (4,1);
  \draw[dashed, line width=0.5pt] (0,-1) -- (4,-1);
  
  \filldraw[black] (0,0) circle node[anchor=east] {0};
  \filldraw[black] (1,0) circle node[anchor=north] {1};
  \filldraw[black] (2.5,0) circle node[anchor=north] {2.5};
  \filldraw[black] (4,0) circle node[anchor=north west] {$T=4$};
  \filldraw[black] (5.5,0) circle node[anchor=north west] {$t$};
  \filldraw[black] (0,1) circle node[anchor=east] {+1};
  \filldraw[black] (0,-1) circle node[anchor=east] {-1};
  \filldraw[black] (0,1.5) circle node[anchor=east] {$h^+$};
  
  \draw[->] (7,0) -- (12.5,0);
  \draw[->] (7,-1.5) -- (7,1.5);
  \draw[line width = 3pt] (11,-1.5) -- (11,1.5);
  \draw[line width = 2pt] (7,-1) -- (8.5,-1);
  \draw[dashed] (8.5,-1) -- (8.5,1);
  \draw[dashed] (9,-1) -- (9,1);
  \filldraw[black] (9,-1) circle (2pt);
  \draw[line width = 2pt] (8.5,1) -- (8.95,1);
  \draw[line width = 2pt] (9.05,1) -- (11,1);
  \draw (9,1) circle (2pt);
  \draw[dashed, line width=0.5pt] (7,1) -- (11,1);
  \draw[dashed, line width=0.5pt] (7,-1) -- (11,-1);
  
  \filldraw[black] (7,0) circle node[anchor=east] {0};
  \filldraw[black] (8.5,0) circle node[anchor=north] {1.5};
  \filldraw[black] (9,0) circle node[anchor=north] {2};
  \filldraw[black] (11,0) circle node[anchor=north west] {$T=4$};
  \filldraw[black] (12.5,0) circle node[anchor=north west] {$t$};
  \filldraw[black] (7,-1) circle node[anchor=east] {-1};
  \filldraw[black] (7,1) circle node[anchor=east] {+1};
  \filldraw[black] (7,1.5) circle node[anchor=east] {$h^-$};
  \node (tl) at (9,1.55) {$\sigma$};
  \draw[->] (tl) -- (9,1.05);
\end{tikzpicture}
    \caption{The illustration of the path segment in $A_{(-1,-1)}^{6,[+,+,-,-,-,+]}(t,[0,1.0,0.5,0.5,0,0.5])$.}
    \label{fig:STEP3FIG2}
\end{figure}
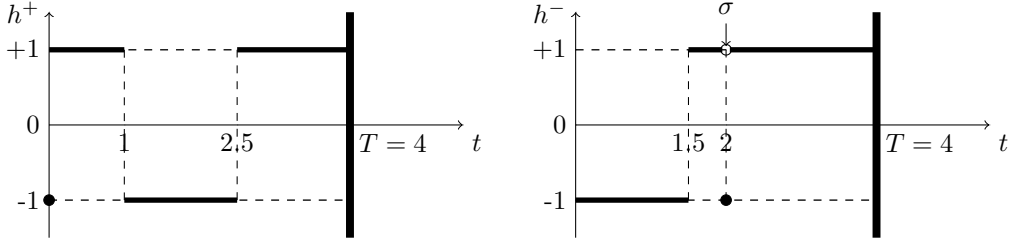
\begin{figure}[!ht]
    \centering
\begin{tikzpicture}
  \draw[->] (0,0) -- (5.5,0);
  \draw[->] (0,-1.5) -- (0,1.5);
  \draw[line width = 3pt] (4,-1.5) -- (4,1.5);
  \filldraw[black] (0,-1) circle (2pt);
  \draw[line width = 2pt] (0,1) -- (1,1);
  \draw[dashed] (1,1) -- (1,-1);
  \draw[line width = 2pt] (1,-1) -- (2.5,-1);
  \draw[dashed] (2.5,-1) -- (2.5,1);
  \draw[line width = 2pt] (2.5,1) -- (4,1);
  \draw[dashed, line width=0.5pt] (0,1) -- (4,1);
  \draw[dashed, line width=0.5pt] (0,-1) -- (4,-1);
  
  \filldraw[black] (0,0) circle node[anchor=east] {0};
  \filldraw[black] (1,0) circle node[anchor=north] {1};
  \filldraw[black] (2.5,0) circle node[anchor=north] {2.5};
  \filldraw[black] (4,0) circle node[anchor=north west] {$T=4$};
  \filldraw[black] (5.5,0) circle node[anchor=north west] {$t$};
  \filldraw[black] (0,1) circle node[anchor=east] {+1};
  \filldraw[black] (0,-1) circle node[anchor=east] {-1};
  \filldraw[black] (0,1.5) circle node[anchor=east] {$h^+$};
  
  \draw[->] (7,0) -- (12.5,0);
  \draw[->] (7,-1.5) -- (7,1.5);
  \draw[line width = 3pt] (11,-1.5) -- (11,1.5);
  \draw[line width = 2pt] (7,-1) -- (8.5,-1);
  \draw[dashed] (8.5,-1) -- (8.5,1);
  \draw[line width = 2pt] (8.5,1) -- (9,1);
  \draw[dashed] (9,1) -- (9,-1);
  \draw[line width = 2pt] (9,-1) -- (11,-1);
  \filldraw[black] (11,+1) circle (3pt);
  \draw[dashed, line width=0.5pt] (7,1) -- (11,1);
  \draw[dashed, line width=0.5pt] (7,-1) -- (11,-1);
  
  \filldraw[black] (7,0) circle node[anchor=east] {0};
  \filldraw[black] (8.5,0) circle node[anchor=north] {1.5};
  \filldraw[black] (9,0) circle node[anchor=north] {2};
  \filldraw[black] (11,0) circle node[anchor=north west] {$T=4$};
  \filldraw[black] (12.5,0) circle node[anchor=north west] {$t$};
  \filldraw[black] (7,-1) circle node[anchor=east] {-1};
  \filldraw[black] (7,1) circle node[anchor=east] {+1};
  \filldraw[black] (7,1.5) circle node[anchor=east] {$h^-$};
  \node (tl) at (11,2) {$\sigma$};
  \draw[->] (tl) -- (11,1.5);
\end{tikzpicture}
    \caption{The illustration of the path segment in $A_{(-1,-1)}^{6,[+,+,-,-,+,-]}(t,[0,1.0,0.5,0.5,0.5,1.5])$.}
    \label{fig:STEP3FIG1}
\end{figure}
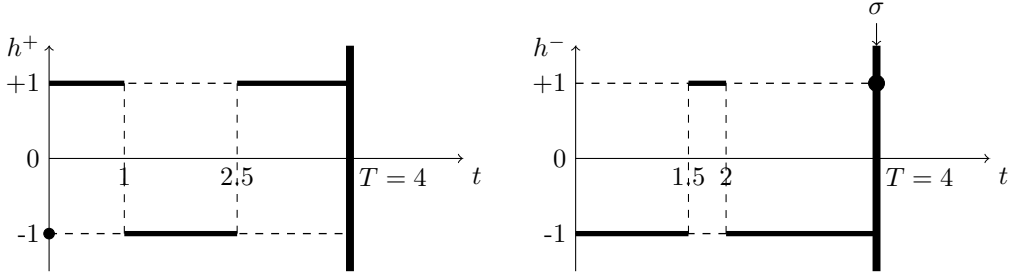

By \Cref{addtwospinflips},
 the value of 
 $A_{(-1,-1)}^{6,[+,+,-,-,-,+]}(t,[0,1.0,0.5,0.5,0,0.5])$ can be directly obtained from
 the value of $A_{(-1,1)}^{4,[+,+,-,+]}(t,[0,1.0,0.5,1.0])$.
In addition, the boundary condition 
 \eqref{eq_boundaryCondition} in \Cref{boundaryCondition}
 allows us to compute the value of 
 $A_{(-1,-1)}^{6,[+,+,-,-,+,-]}(t,[0,1.0,0.5,0.5,0.5,1.5])$
 by the value of
 $A_{(-1,-1)}^{5,[+,+,-,-,+]}(t,[0,1.0,0.5,0.5,0.5])$. In summary, our final estimation is
\begin{equation*}
\begin{split}
     &A_{(-1,-1)}^{6,[+,+,-,-,+,-]}(t,[0,1.0,0.5,0.5,0.5,1.0]) \\
   \approx &-\frac{1}{4}
   \mel{+1}{H_0}{-1}\mel{-1}{H_0}{+1} 
   A_{(-1,-1)}^{4,[+,+,-,+]}(t,[0,1.0,0.5,1.0])\\
   &+\frac{3}{4} \ii
   \mel{-1}{H_0}{+1}
   A_{(-1,-1)}^{5,[+,+,-,-,+]}(t,[0,1.0,0.5,0.5,0.5]).
\end{split}
\end{equation*}
Therefore, we estimate the value of $A_{(-1,-1)}^{6,[+,+,-,-,+,-]}(t,[0,1.0,0.5,0.5,0.5,1.0])$
 from those with $D\leqslant D_{\max}$.

The idea can be generalized to any paths with $D_{\max} + 1$ spin flips in a straightforward way.
In the example above, we choose to shift the last spin flip of negative branch. 
Clearly, the same approach also works for the positive branch.
When the negative branch is chosen, $\sigma$ can take values in the interval $[2,4]$, which has a length of $2$. 
If the positive branch is chosen, 
 the corresponding interval will have a length of $3$. 
In this case, we prefer to choose the negative branch since the interpolation on a shorter interval is more likely to get a better approximation.
Technically, to determine which branch to choose,
 we just need to look at the last three signs of $\sgn$.
If at least two of them are positive/negative, we choose the positive/negative branch to formulate the closure. This approach applies to any case with $D_{\max} \geqslant 2$.

In general, when a path segment $h$ has $D$ spin flips with $D = D_{\max} + 1$, after selecting the branch by the aforementioned method, we approximate $A(t,h)$ by linear interpolation.
In the following discussion,
 we assume that the positive branch is selected.
The procedure is the same if the negative branch is selected.
According to the branch selection, there must be at least two spin flips on the chosen branch ($h^+$ according to our assumption).
Let $\sigma_0$ and $\sigma_*$ be the locations of the last two spin flips on $h^+$, and $\sigma_* < \sigma_0$. 
If we move the last spin flip on $h^+$ to a different location $\sigma \in (\sigma_*, T)$, we obtain
a new path $h_\sigma=(h_\sigma^+,h_\sigma^-):[0,T)\rightarrow\mathcal{S}^2$ defined by
\begin{equation*}
    \begin{split}
        h_\sigma^+(\tau) &= \begin{cases}
        h^+(\tau), & \text{ if } \tau < \sigma^* \\
        h^+(\sigma_*), & \text{ if } \sigma_* \leqslant \tau < \sigma \\
        h^+(T^-), & \text{ if } \sigma \leqslant \tau < T
        \end{cases}.\\
        h_\sigma^-(\tau) &= h^-(\tau). 
    \end{split}
\end{equation*}
where $h^+(T^-) = \lim_{s\rightarrow T^-} h^+(s)$.
Clearly, we have $h = h_{\sigma_0}$
 and thus we can approximate $A(t,h)$ by the following linear interpolation:
\begin{equation}
    A(t,h) = A(t,h_{\sigma_0}) \approx \frac{T - \sigma_0}{T - \sigma_*} A(t,h_{\sigma_*})
    + \frac{\sigma_0 - \sigma_*}{T-\sigma_*} A(t,h_T).
    \label{interpolation}
\end{equation}
According to \Cref{boundaryCondition} and \Cref{addtwospinflips}, we have
\begin{equation}
    A(t,h_T) = -\ii A(t,h') \mel{\hat{h}^+(T^-)}{H_0}{h^+(T^-)}, \qquad
    A(t,h_{{\sigma}_*}) = - A(t,h'')\mel{+1}{H_0}{-1}\mel{-1}{H_0}{+1}.
    \label{nodevalues}
\end{equation}
where the paths $h'=(h'^+,h'^-)$ and $h''=(h''^+,h''^-)$ are given by
\begin{align*}
   h'^+(\tau) &= \begin{cases}
   h^+(\tau), & \text{ if } \tau < \sigma^*, \\
   h^+(\sigma_*), & \text{ if } \sigma_* \leqslant \tau < T,
   \end{cases} & h''^+(\tau) &= \begin{cases}
   h^+(\tau), & \text{ if } \tau < \sigma^*, \\
   \lim_{s\rightarrow \sigma_*^-} h^+(s), & \text{ if } \sigma_* \leqslant \tau < T,
   \end{cases}\\
   h'^-(\tau) &= h^-(\tau), &
   h''^-(\tau) &= h^-(\tau). 
\end{align*}
It can be seen that $h'$ has one less spin flip than $h$, and $h''$ has two less spin flips than $h$. Therefore both $A(t,h')$ and $A(t,h'')$ are included in the truncated system. Plugging \eqref{nodevalues} into \eqref{interpolation} yields an estimated value of $A(t,h)$, which forms the closure of the PDE system \eqref{PDEoriginal}.

In the next section, we will find the observable of the system by solve the PDE system numerically. We call this approach differential equation based path integral (DEBPI) method.

\begin{remark}
Here we would like to comment on the relationship between the i-QuAPI scheme and the PDE system \eqref{PDEoriginal}. According to the
 the iteration \eqref{propagation}, i-QuAPI considers $2^{2\dk}$ equations in the system.
However, the i-QuAPI method does not perform the truncation by simply choosing a maximum value of $D$. Instead, it considers path segments with at most $\dk$ spin flips on each branch, so that the maximum spin flips in the path segment can reach $2\dk$, but not all path segments with $2\dk$ spin flips are taken into account.
Besides, the treatment of both boundary condition and closure of the system can be observed from \eqref{propagation}.
The boundary condition is needed when updating the value of $A$ for path segments with a spin flip near the end of the path, which corresponds to the case $S_k \neq S_{k-1}$. In this case, such a spin flip is simply dropped on the right-hand side of \eqref{propagation}, which is similar to the right-hand side of our boundary condition \eqref{eq_boundaryCondition}. The extra bra-ket terms in \eqref{eq_boundaryCondition} appear in the i-QuAPI method as the bra-ket terms in $I(S_k,S_{k-1})$ defined in \eqref{definitionI}. As for the closure, according to the discretization of the i-QuAPI method, the closure is required only when there are $\dk$ spin flips on either of the branches, and in this case, we have $\tau_1 = \dt$ and $\tau_1 + \cdots + \tau_D = T-\dt$. According to \eqref{PDEoriginal}, one needs the values of $A_{(r^+,r^-)}^{D+1,[\pm,\sgn]}(t,[0, \tau_1, \cdots, \tau_D])$ to close the system. From this point of view, the approach of i-QuAPI is to assume that
\begin{displaymath}
\begin{split}
& A_{(r^+,r^-)}^{D+1,[\pm,\sgn]}(t,[0, \tau_1, \cdots, \tau_D]) = A_{(r^+,r^-)}^{D+1,[\pm,\sgn]}(t,[\dt, \tau_1, \cdots, \tau_D]) \\
={} & A_{(r^+,r^-)}^{D,[\pm,\sgn_{1:D-1}]}(t,[\dt, \tau_1, \cdots, \tau_{D-1}])
\left(-\ii\mel{r_f^+}{H_0}{\hat{r}_f^+}
\delta_{\sgn_D}^+
+ \ii \mel{\hat{r}_f^-}{H_0}{r_f^-}
\delta_{\sgn_D}^-\right),
\end{split}
\end{displaymath}
where the boundary condition \eqref{eq_boundaryCondition} has been applied. Based on such a system closure, the i-QuAPI method can be considered as a numerical scheme of \eqref{PDEoriginal}.
\end{remark}


\section{Numerical Experiments}
\label{Sec_Numerical_Experiments}
To verify the correctness of the DEBPI method and show the possible memory savings by solving \eqref{PDEoriginal} directly, we present several numerical experiments in this section. The numerical method to solve \eqref{PDEoriginal} generally follows the standard numerical solvers of hyperbolic systems, which will be briefly introduced in \Cref{sec:num_method}. The numerical results will be exhibited and discussed in \Cref{sec:num_res}.

\subsection{Numerical Method}
\label{sec:num_method}
The general structure of our numerical scheme is based on the Strang splitting method to deal with the zeroth- and first-order terms on the right-hand side of \eqref{PDEoriginal} separately. 
In general, for a differential equation having the form
\begin{equation*}
    \frac{\partial}{\partial t} u = L_1 u + L_2 u
\end{equation*}
 with $L_1$ and $L_2$ being two linear operators, Strang splitting approximates the solution by
\begin{equation*}
    u(t+\dt) \approx \e^{L_1 \dt/2} \e^{L_2 \dt} \e^{L_1 \dt/2} u(t).
\end{equation*}
In our case, we need to focus on the numerical solver of the following two equations:
\begin{align}
        \frac{\partial}{\partial t} A_{(s^+,s^-)}^{D,\sgn}(t,[\tau_1,\cdots,\tau_D])
        =& \frac{\partial}{\partial \tau_1} 
        A_{(s^+,s^-)}^{D,\sgn}(t,[\tau_1,\cdots,\tau_D]);
        \label{eq_advection}\\
    \begin{split}
        \frac{\partial}{\partial t} A_{(s^+,s^-)}^{D,\sgn}(t,[\tau_1,\cdots,\tau_D])
        =& -W_{(s^+,s^-)}^{D,\sgn}\left([\tau_1,\cdots,\tau_D]\right)
		A_{(s^+,s^-)}^{D,\sgn}\left(t,[\tau_1,\cdots,\tau_D]\right) \\
		&+ A_{(s^+,\hat{s}^-)}^{D+1,[-,\sgn]}
		\left(t,[0,\tau_1,\cdots,\tau_D]\right)
		+ A_{(\hat{s}^+,s^-)}^{D+1,[+,\sgn]}
		\left(t,[0,\tau_1,\cdots,\tau_D]\right).
    \end{split}
    \label{eq_source}
\end{align}
Thus, evolving \eqref{PDEoriginal} by one time step $\Delta t$ consists of solving \eqref{eq_advection} by $\dt/2$, and then solving \eqref{eq_source} by $\Delta t$, followed by solving \eqref{eq_advection} again by $\Delta t /2$.

\begin{figure}[ht]
    \centering
    \subfloat[$D=2$, $N=5$]{
    \label{fig:2d}
\begin{tikzpicture}
    \draw[->] (-0.5,0) -- (5.5,0);
    \draw[->] (0,-0.5) -- (0,5.5);
    \draw[line width = 1.5pt] (0,0) -- (5,0) -- (0,5) -- cycle;
    \filldraw[black] (5.5,0) circle node[anchor=north] {$\tau_1$};
    \filldraw[black] (0,5.5) circle node[anchor=east] {$\tau_2$};
    \filldraw[black] (5,0) circle node[anchor=north] {$T$};
    \filldraw[black] (0,5) circle node[anchor=east] {$T$};
    \filldraw[black] (0,0) circle node[anchor=north east] {$0$};
    \filldraw[red] (0,0) circle (2pt);
    \filldraw[red] (1,0) circle (2pt);
    \filldraw[red] (2,0) circle (2pt);
    \filldraw[red] (3,0) circle (2pt);
    \filldraw[red] (4,0) circle (2pt);
    \filldraw[red] (0,1) circle (2pt);
    \filldraw[red] (1,1) circle (2pt);
    \filldraw[red] (2,1) circle (2pt);
    \filldraw[red] (3,1) circle (2pt);
    \filldraw[red] (0,2) circle (2pt);
    \filldraw[red] (1,2) circle (2pt);
    \filldraw[red] (2,2) circle (2pt);
    \filldraw[red] (0,3) circle (2pt);
    \filldraw[red] (1,3) circle (2pt);
    \filldraw[red] (0,4) circle (2pt);
\end{tikzpicture}}
\qquad
    \subfloat[$D=3$, $N=4$]{
    \label{fig:3d}
\begin{tikzpicture}
    \draw[->] (9,2) -- (12.5,2);
    \draw[->] (9,2) -- (9,5.5);
    \draw[->] (9,2) -- (7.3,0);
    \draw[line width = 1.5pt] (9,2) -- (12,2) -- (9,5) -- cycle;
    \draw[line width = 1.5pt] (9,2) -- (7.725,0.5) -- (12,2);
    \draw[line width = 1.5pt] (9,5) -- (7.725,0.5);
    \filldraw[black] (7.3,0) circle node[anchor=north] {$\tau_1$};
    \filldraw[black] (12.5,2) circle node[anchor=north] {$\tau_2$};
    \filldraw[black] (9,5.5) circle node[anchor=east] {$\tau_3$};
    \filldraw[black] (12,2) circle node[anchor=north] {$T$};
    \filldraw[black] (9,5) circle node[anchor=east] {$T$};
    \filldraw[black] (7.725,0.5) circle node[anchor=east] {$T$};
    
    \filldraw[red] (9,2) circle (2pt);
    \filldraw[red] (9,2.75) circle (2pt);
    \filldraw[red] (9,3.5) circle (2pt);
    \filldraw[red] (9,4.25) circle (2pt);
    \filldraw[red] (9.75,2) circle (2pt);
    \filldraw[red] (9.75,2.75) circle (2pt);
    \filldraw[red] (9.75,3.5) circle (2pt);
    \filldraw[red] (10.5,2) circle (2pt);
    \filldraw[red] (10.5,2.75) circle (2pt);
    \filldraw[red] (11.25,2) circle (2pt);

    \filldraw[blue] (9-0.31875,2-0.375) circle (2pt);
    \filldraw[blue] (9-0.31875,2.75-0.375) circle (2pt);
    \filldraw[blue] (9-0.31875,3.5-0.375) circle (2pt);
    \filldraw[blue] (9.75-0.31875,2-0.375) circle (2pt);
    \filldraw[blue] (9.75-0.31875,2.75-0.375) circle (2pt);
    \filldraw[blue] (10.5-0.31875,2-0.375) circle (2pt);
    
    \filldraw[green] (9-2*0.31875,2-2*0.375) circle (2pt);
    \filldraw[green] (9-2*0.31875,2.75-2*0.375) circle (2pt);
    \filldraw[green] (9.75-2*0.31875,2-2*0.375) circle (2pt);
    
    \filldraw[purple] (9-3*0.31875,2-3*0.375) circle (2pt);
        
\end{tikzpicture}}
    \caption{The grid points on the two-dimensional simplex (red points) with $N=5$ and three-dimensional simplex (points of all colors) with $N=4$.}
    \label{fig:demonstrationOfGridPoints}
\end{figure}
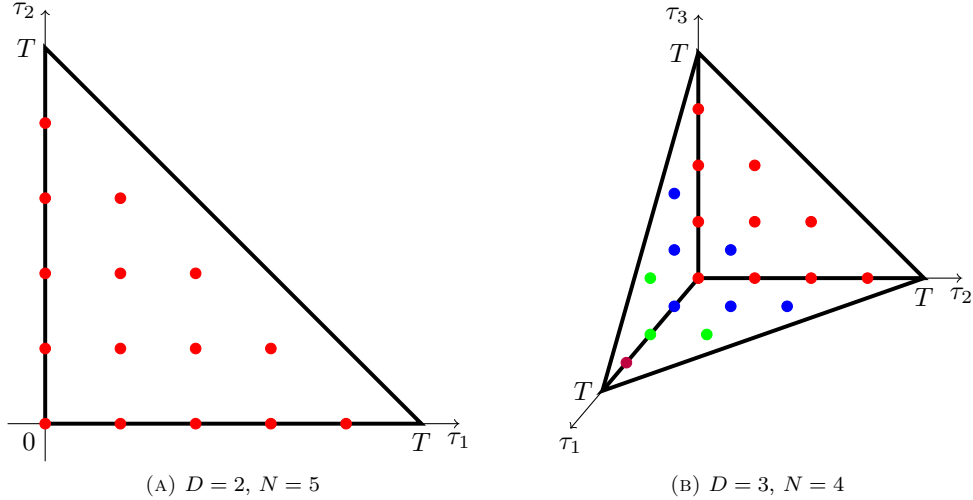

For both equations, a uniform mesh is used to discretize the domain $\overline{\triangle_T^{(D)}}$. Note that this is not needed for $D = 0$ since the path segment $h$ is fully determined by its initial state. When $D > 0$, the sets of grid points are
\begin{equation*}
    \mathcal{P}_D = \left\{ \left(\frac{m_1h_s}{N},\cdots,\frac{m_Dh_s}{N}\right); m_k \in \mathbb{Z}^+, \sum_{k=1}^D m_k < N \right\},
    \quad D = 1,\cdots,D_{\max}.
\end{equation*}
where $h_s = T/N$ is the grid size.
For example, \Cref{fig:demonstrationOfGridPoints}
 shows the grid points in the case of $D = 2, N=5$ and $D = 3, N=4$.
We do not place grid points on the hypotenuse since
 their values are determined
 by the boundary condition
 as discussed in \Cref{boundaryCondition}.

The $\tau_1$-derivative in \eqref{eq_advection} is discretized by the finite difference method.
As \eqref{eq_advection} is a hyperbolic equation with linear advection,
 we apply second-order upwind scheme 
 and Lax-Wendroff method
 for the $\tau_1$-derivative.
Although there are no grid points on the hypotenuse,
 we still can calculate the values of $A$ on the hypotenuse 
 with \Cref{boundaryCondition}.
Therefore, the $\tau_1$-derivative of all points 
 with $\tau_1+\cdots+\tau_D\leqslant N-2$ 
 can be estimated by second-order upwind scheme.
As for the grid points with $\tau_1+\cdots+\tau_D=N-1$,
 we apply Lax-Wendroff method for the estimation of $\tau_1$-derivative.
\eqref{eq_source} can be regarded as an ODE with respect to $t$.
Therefore, we simply employ Runge-Kutta method for the numerical solution of \eqref{eq_source}.
Note that the estimation for the terms with $D_{\max}+1$ spin flips 
 is discussed in \Cref{ClosureOfTheSystem}.


As a summary, we provide the outline of our algorithm below:
\begin{itemize}
    \item For each set of parameters
    $D = 0,\cdots,D_{\max}$, $(s^+,s^-) = (\pm1,\pm1)$
    and $p\in\mathcal{P}_D$
    compute 
    $A_{(s^+,s^-)}^{D,\sgn}(0,p)$
    according to \eqref{Asimplified}.
    \item Assume that we already have all values of $A_{(s^+,s^-)}^{D,\sgn}(t,p)$ for $D = 0,\cdots,D_{\max}$
    all possible choices of $(s^+,s^-)$ and $p\in\mathcal{P}_D$
    at specific time $t$.
    We evolve the system to obtain corresponding values of at time $t+\dt$ by the following steps.
    \begin{itemize}
        \item Evolve the system \eqref{eq_advection} by a half time step $\dt/2$ using the second-order upwind scheme (for the points with $\tau_1+\cdots+\tau_D\leqslant N-2$) or Lax-Wendroff method (for the points with $\tau_1+\cdots+\tau_D = N-1$).
        If points on hypotenuse (the sum of all components are $T$) are involved,
            boundary condition \eqref{eq_boundaryCondition} is applied.
        \item Evolve the system \eqref{eq_source} by a full time step $\dt$.
        If points on the $(D_{\max}+1)$-dimensional simplex is involved,
        the method to close the system in \Cref{ClosureOfTheSystem}
         is applied.
        \item Evolve the system \eqref{eq_advection} by another half time step $\dt/2$ using the second-order upwind scheme or Lax-Wendroff method again.
        \item Compute the density matrix at time $t+\dt$ according to \eqref{densityMatrixContinuous} and calculate the observable
        by $\expval{O(t)} = \tr(O \rho_s(t))$.
    \end{itemize}
    \item Repeat the previous step to evolve the system iteratively.
\end{itemize}
Here the first step prepares the initial data, which requires  numerical integration on $\overline{\triangle_T^{(D)}}$.
For the evolution of numerical solution, we fix the time step $\dt$ so that the solvers of both equations
\eqref{eq_advection}\eqref{eq_source} can be written as matrix-vector multiplications, and the matrices are the same for all steps.

At the end of this subsection, we would like to discuss the memory cost of our method.
For a $D$-dimensional simplex with $N$ nodes on each edge,
 there are $\begin{pmatrix}N+D\\D\end{pmatrix}$ grid points in one simplex.
Besides, the initial state of $h$, \emph{i.e.} the subscript $(r^+, r^-)$ of $A$, has four different possibilities for the spin-boson model; the branches of the spin flips, \emph{i.e.} $\sgn$ in the superscript of $A$, can take $2^D$ choices.
Therefore, in total, we need
\begin{equation} \label{eq:ndof}
N_{\mathrm{dof}} = 4\sum_{D=0}^{D_{\max}} 2^D \begin{pmatrix} N+D \\ D \end{pmatrix}
\end{equation}
degrees of freedom to store the numerical solution.
To compare this with the i-QuAPI method,
 we first recall that i-QuAPI needs to store all values of the map $A_l:\mathcal{S}^{2\dk}\rightarrow \mathbb{C}$
 in each iteration, so the memory cost is $2^{2\dk}$.
Assume that both methods have same truncation time and grid size, we then have $\dk=N$.
Thus we need to compare $N_{\mathrm{dof}}$ and $2^{2N}$. To this end, we estimate $N_{\mathrm{dof}}$ using Stirling's approximation $k!\sim \sqrt{2\pi}\e^{-k} k^{k+1/2}$:
\begin{equation*}
    \begin{split}
    N_{\mathrm{dof}} &< 4 (D_{\max}+1) 2^{D_{\max}} \begin{pmatrix}N+D_{\max}\\ D_{\max}\end{pmatrix} \\
    &\lesssim \frac{4 (D_{\max}+1) 2^{D_{\max}}}{D_{\max}!}
    \frac{(N+D_{\max})^{N+D_{\max}+1/2}}
    {\e^{D_{\max}}N^{N+1/2}} \\
    &\lesssim \frac{4(D_{\max}+1)2^{D_{\max}}}{D_{\max}!} (N+D_{\max})^{D_{\max}}.
    \end{split}
\end{equation*}
It is now clear that if $D_{\max} \ll N$, the value of $N_{\mathrm{dof}}$ can be significantly less than $2^{2N}$
so that our DEPBI method will require less memory than the i-QuAPI method.
 
\subsection{Numerical Results}
\label{sec:num_res}
In all our experiments,
 the frequencies $\omega_j$ are distributed in $[0,\omega_{\max}]$ following the Poisson distribution:
\begin{equation*}
    \omega_j = - \omega_c \ln \left(
    1 - \frac{j}{L}(1-\exp(-\omega_{\max}/\omega_c))\right),
    \quad j=1,\cdots,L
\end{equation*}
where $L$ is the number of harmonic oscillators.
The coupling intensity $c_j$ is
\begin{equation*}
    c_j = \omega_j \sqrt{\frac{\xi\omega_c}{L}(1-\exp(-\omega_{\max}/\omega_c))},
    \quad j=1,\cdots,L.
\end{equation*}
These parameters correspond to the bath with Ohmic spectral density \cite{makri1999linear}.
The number of harmonic oscillators $L$ is chosen to be $200$, and the maximum frequency is chosen as $\omega_{\max} = 4\omega_c$. For the DEPBI method, the time step $\dt$ is fixed as $1/80$ in all cases. Other parameters will be specified for each experiment.

\subsubsection{Experiments with different coupling intensities}
\begin{figure}[ht]
    \subfloat[$\xi=0.2$]{\includegraphics[width= 0.45\textwidth]{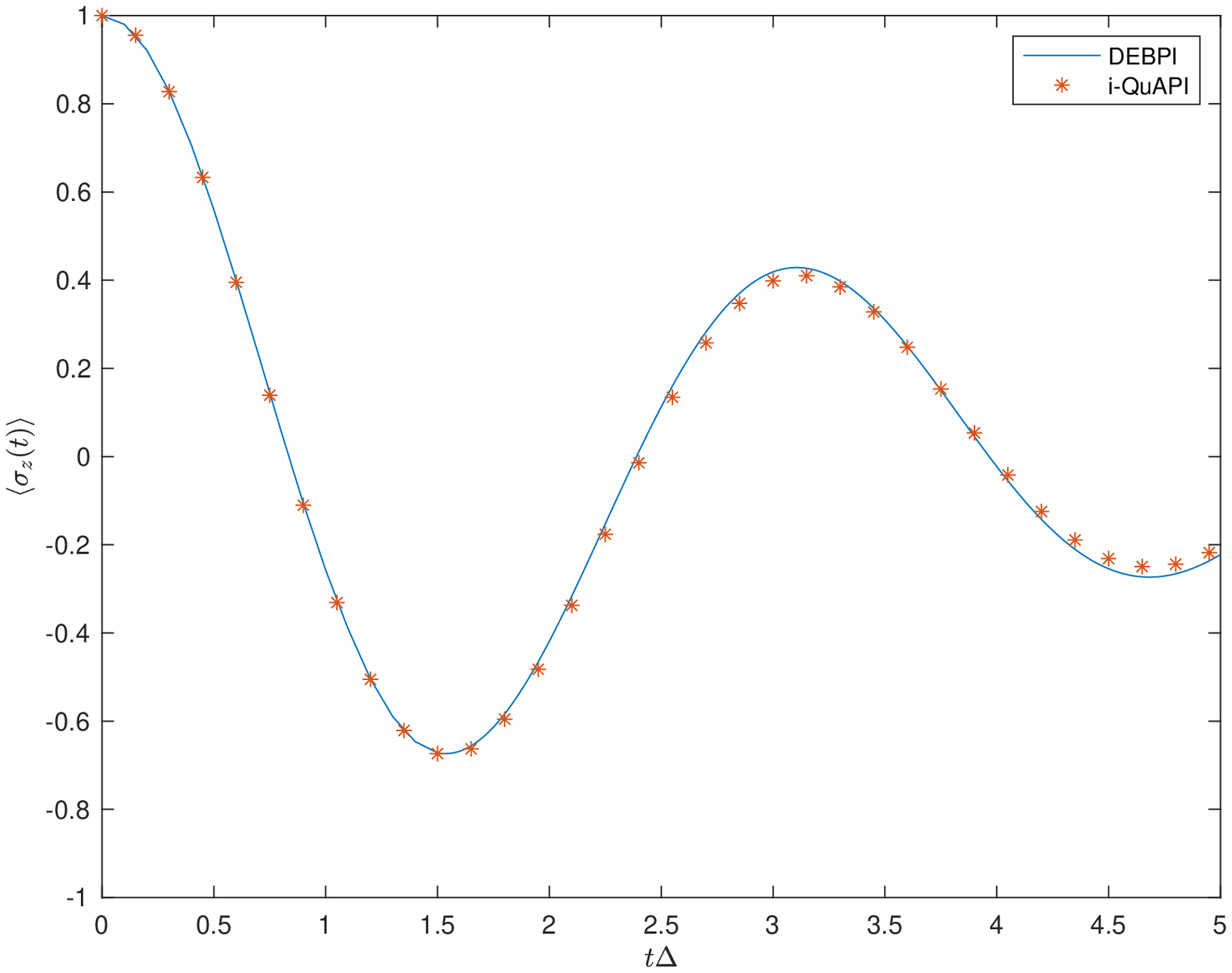}}\qquad
    \subfloat[$\xi=0.4$]{\includegraphics[width= 0.45\textwidth]{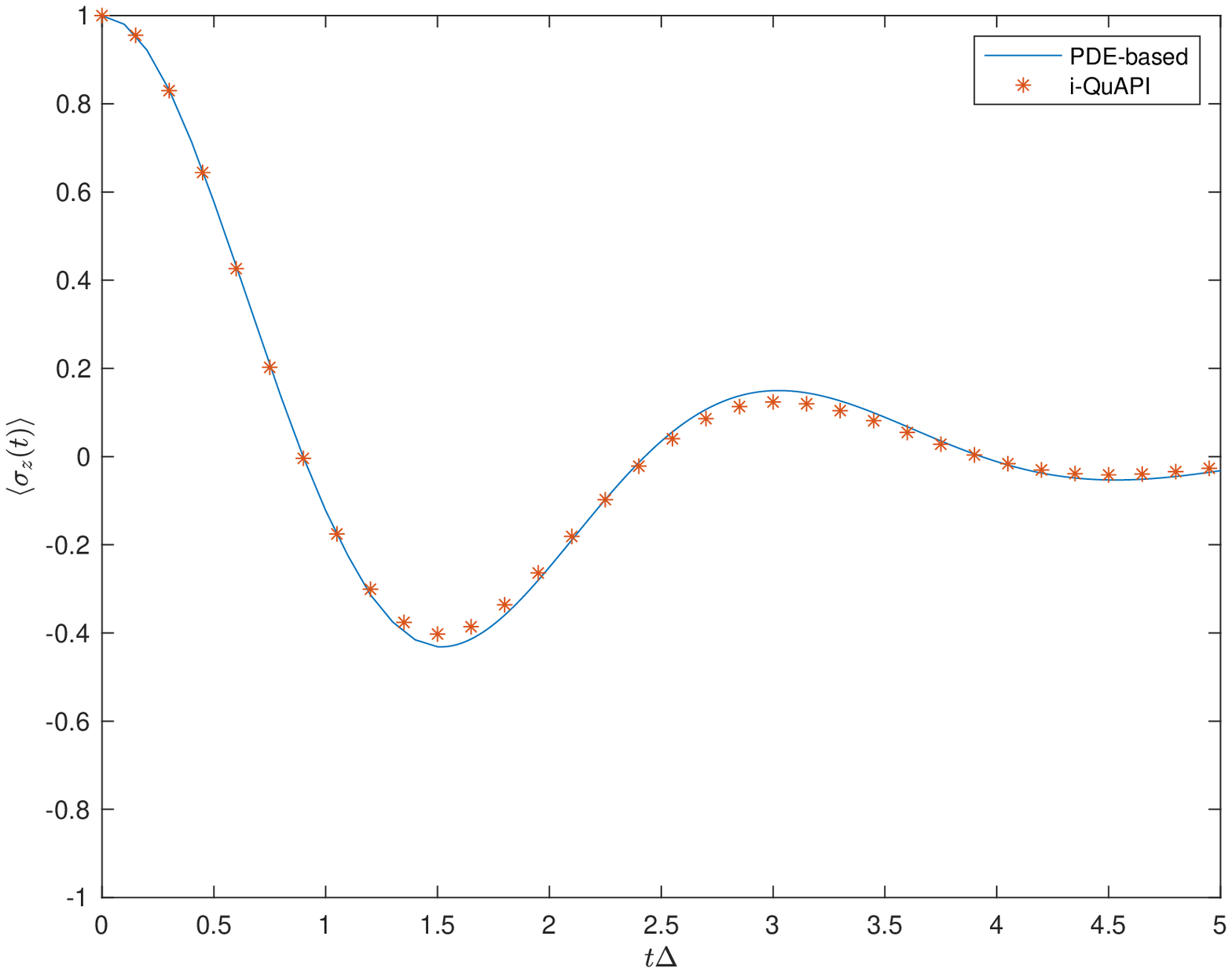}}
    \caption{Both methods use truncation time $T=1.5$. For i-QuAPI method, $\dk$ is chosen to be 10. For c-QuAPI, $D_{\max}=8$ and $N=8$.}
    \label{fig_result_data0113}
\end{figure}

In order to check the validity of our method, we first study the following parameters, which have been considered in \cite{Kelly2013efficient,cai2020inchworm}:
\begin{displaymath}
\Delta=1, \quad \omega_c = 2.5\Delta, \quad \beta = 5/\Delta, \quad \epsilon=0.
\end{displaymath}
The numerical results for $\xi = 0.2$ and $0.4$ are given in \Cref{fig_result_data0113}. The results of our method is compared with the i-QuAPI results, and the parameters used for both methods are given in the caption. The results correctly show that when the bath-system coupling is stronger, the fluctuation of the observable damps faster. In both cases, the evolution of the observable matches well with each other, showing reliability of our approach. 

According to \eqref{eq:ndof}, it can be calculated that the DEBPI method needs to store $N_{\mathrm{dof}} = 17444860$ different values of $A(t,h)$ for each time step.
In comparison, the i-QuAPI method, which has a smaller grid size, only requires to store $2^{2\times 10} = 1048576$ different values. To understand such a difference, we note that each discrete path segment in the i-QuAPI method can also be represented using the initial state $(r^+, r^-)$, the number of spin flips $D$, the branches of the spin flips $\sgn$, and the time difference between each pair of spin flips $(\tau_1, \cdots, \tau_D)$. However, for the i-QuAPI method, there are never more than one spin flip occurring at the same time on the same branch. In other words, if $\sgn_k = \sgn_{k+1}$, then $\tau_{k+1}$ must be nonzero. For example, when $D = 2$ and $\sgn_1 = \sgn_2$, the five points located on the vertical axis of \Cref{fig:2d} are not considered in the i-QuAPI method, leading to a lower memory cost than that in our discretization. For larger $D$, this will cause more significant difference in the memory usage even if the grid sizes for both methods are the same. This difference can be eliminated by using a smarter PDE solver, which will be studied to our future work.

Based on our current approach, the memory cost grows exponentially with $D_{\max}$. Therefore, if the problem setting allows a lower value of $D_{\max}$, the DEBPI method will show its advantage. Moreover, if $D_{\max}$ can be chosen small, we can use a small grid size in the discretization of $\triangle_T^{(D)}$ to achieve better accuracy, while for the i-QuAPI method, the memory cost grows exponentially with $\dk$, so that it is prohibitive to choose small grid sizes. In the following two subsections, we will show several cases with small spin-flipping frequency such that our method can achieve a lower memory cost.

\subsubsection{Experiments with different biases} \label{sec:bias}
\begin{figure}[ht]
    \subfloat[$\epsilon=0$]{\includegraphics[width= 0.3\textwidth]{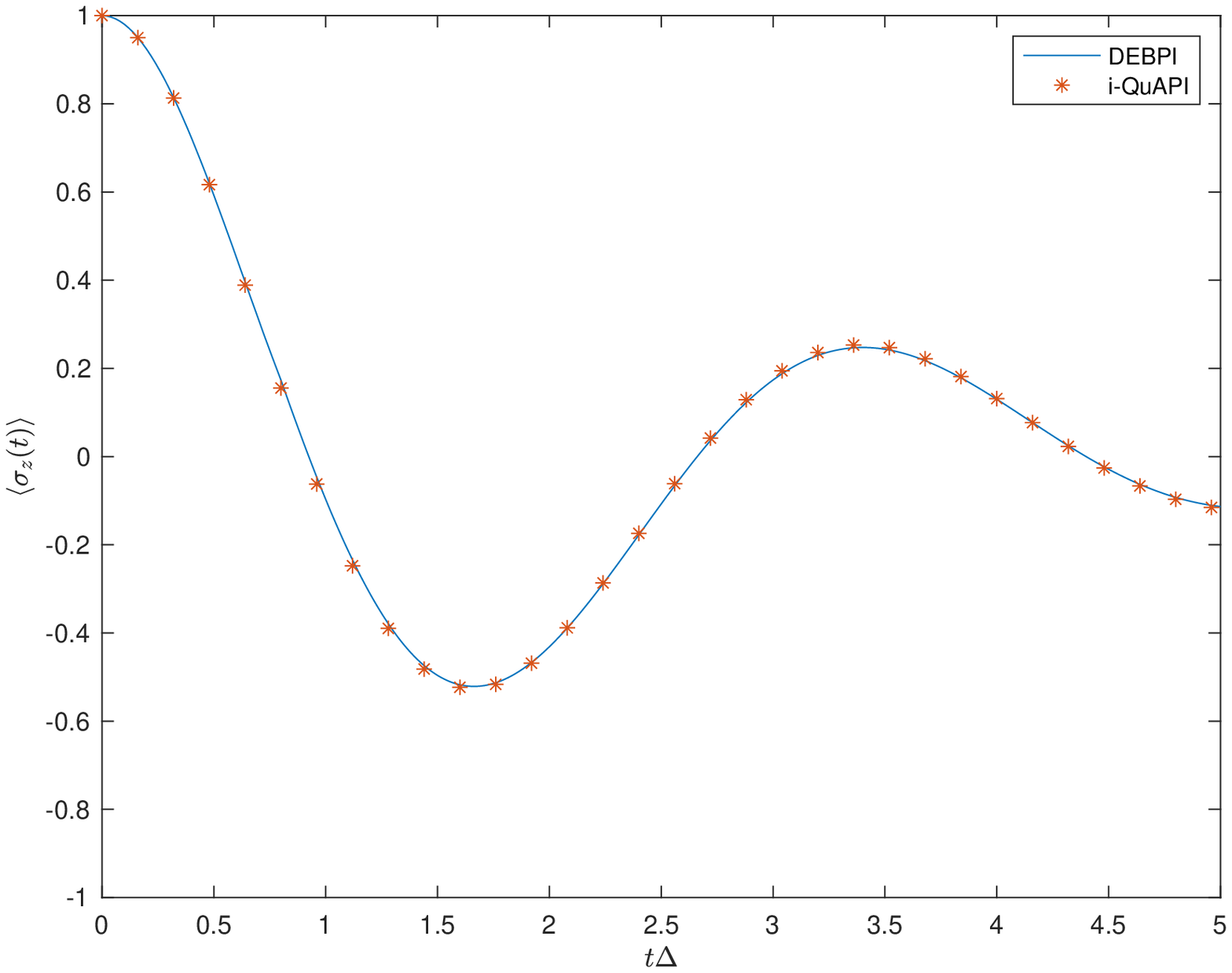}} \quad
    \subfloat[$\epsilon=0.5\Delta$]{\includegraphics[width= 0.3\textwidth]{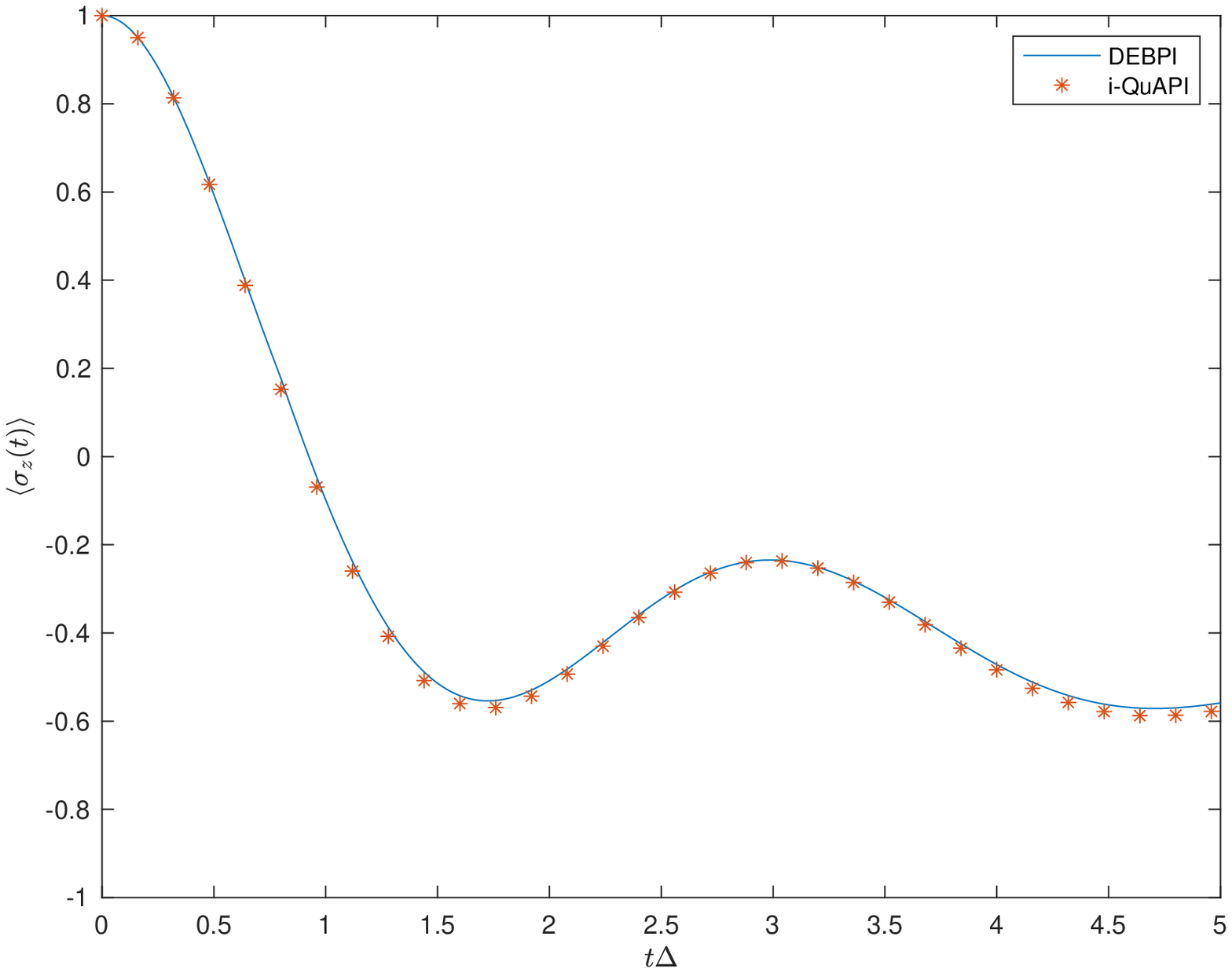}} \quad
    \subfloat[$\epsilon=\Delta$]{\includegraphics[width= 0.3\textwidth]{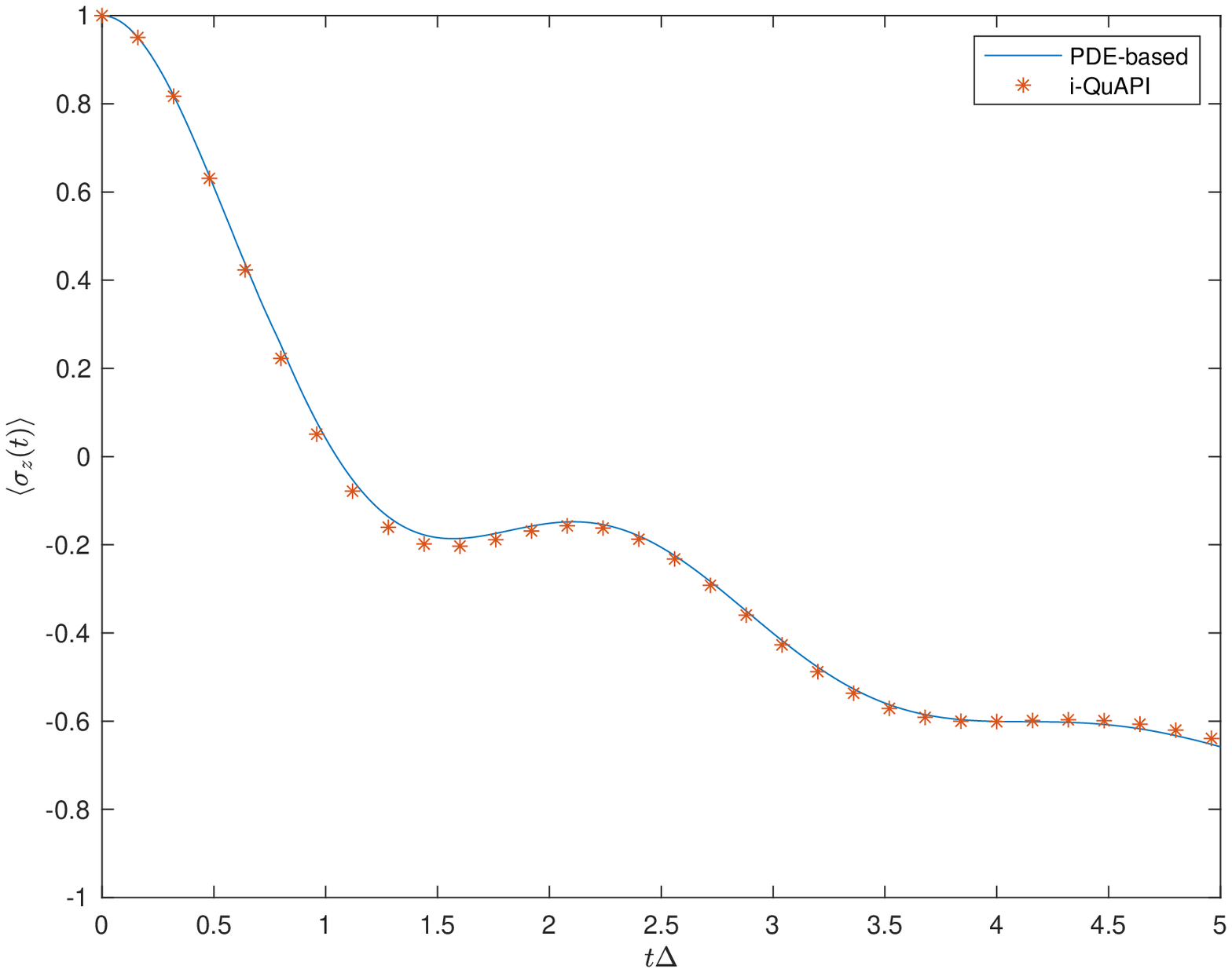}}
    \caption{Both methods use truncation time $T=4$. For i-QuAPI method, $\dk$ is chosen to be 10 and for c-QuAPI, $D_{\max}=5$ and $N=10$.}
    \label{fig_result_data101112}
\end{figure}

We now consider the following set of parameters:
\begin{displaymath}
 \xi = 0.2, \quad \Delta=0.2, \quad \omega_c = 5\Delta, \quad \beta = 5/\Delta,
\end{displaymath}
where the spin flipping frequency $\Delta$ is smaller so we expect that a smaller $D_{\max}$ can be adopted for our DEBPI approach. Three different biases $\epsilon$ are chosen, and the results are given in
 \Cref{fig_result_data101112}. For $D_{\max} = 5$, the curves for the DEPBI method almost coincide with the i-QuAPI results. Since the parameter $\epsilon$ denotes the difference between the energies of the two states, the three figures correctly show that when $\epsilon$ is larger, the state of spin is more likely to be found as $\ket{-1}$, leading to the downshift of the curve in the second and third subplots in \Cref{fig_result_data101112}.

\begin{figure}
    \centering
    \includegraphics[width=0.5\textwidth]{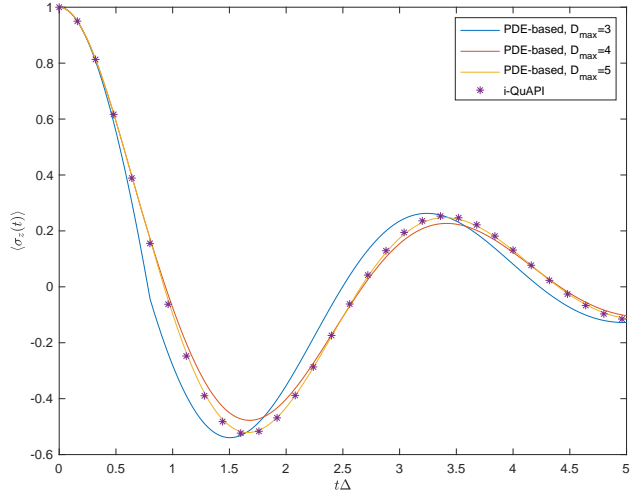}
    \caption{$\epsilon=0$ and $D_{\max}=3,4,5$.}
    \label{differentDmax}
\end{figure}

In these tests, DEBPI needs
 $\displaystyle\sum_{D=0}^5 4\times 2^D \begin{pmatrix} D+10 \\ D \end{pmatrix} = 458748$ numbers to store the numerical solution, while
 the i-QuAPI scheme needs $2^{2\times 10} = 1048576$. Thus the memory cost of our method is about a half of the cost of i-QuAPI. Note that further reducing $D_{\max}$ may cause significant error in the numerical solution.
\Cref{differentDmax} shows the numerical results
 for $D_{\max} = 3, 4, 5$ when $\epsilon = 0$, indicating that $D_{\max} = 5$ is a proper choice to guarantee the quality of the solution.
 
In general, the parameter $D_{\max}$ can be considered as a factor controlling the trade-off between the memory cost and the numerical accuracy.
In order that $D_{\max}$ can be chosen small, we need both the memory time and the spin flipping frequency to be relatively small, meaning that within the path segment $T$, 
the spin is unlikely to flip many times.
Under such circumstances, our method is more likely to outperform the method of i-QuAPI in terms of the memory cost.
 
\subsubsection{Experiments with different temperatures}
\begin{figure}[ht]
    \subfloat[$\beta=0.2/\Delta$]{\includegraphics[width= 0.3\textwidth]{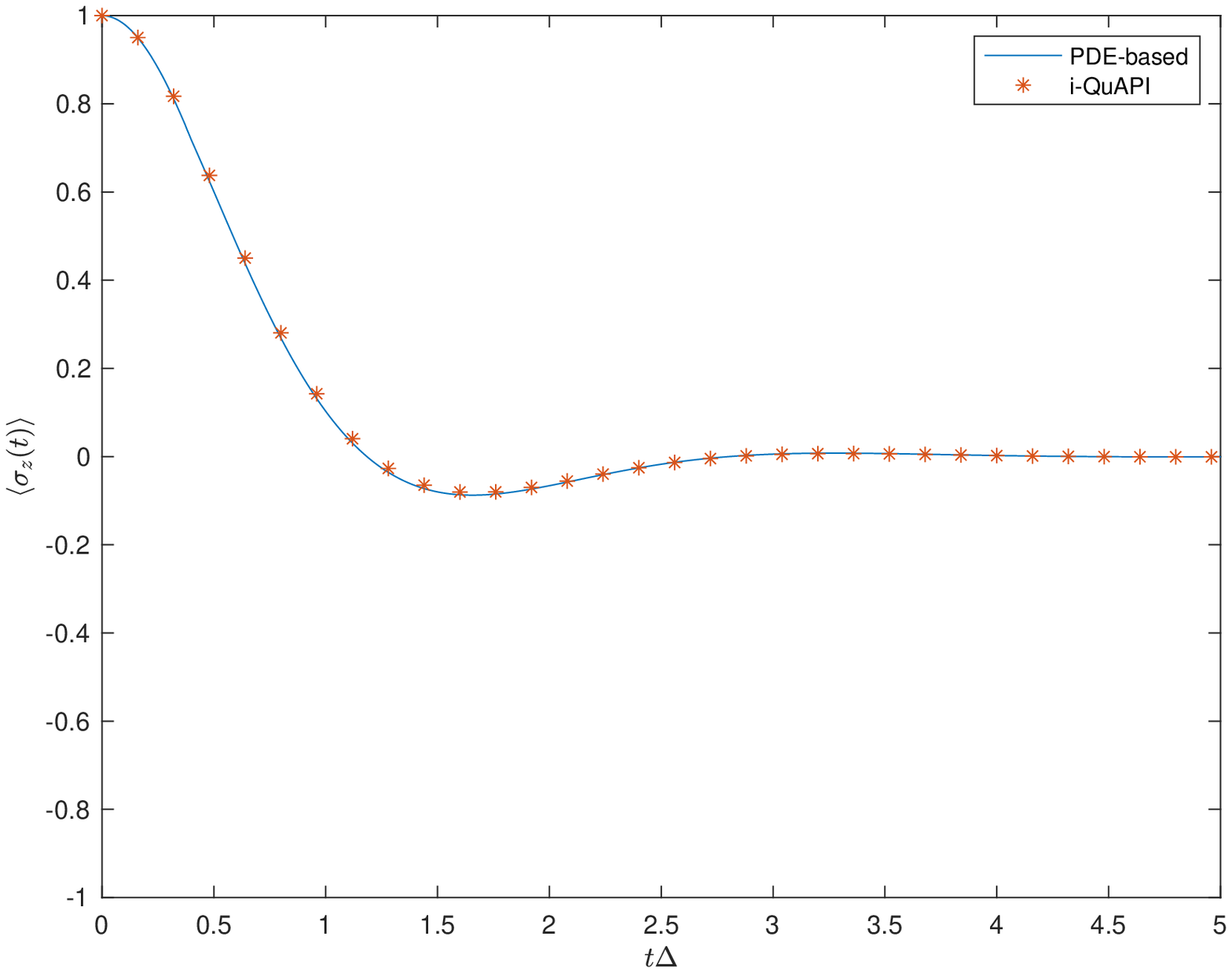}} \quad
    \subfloat[$\beta=1/\Delta$]{\includegraphics[width= 0.3\textwidth]{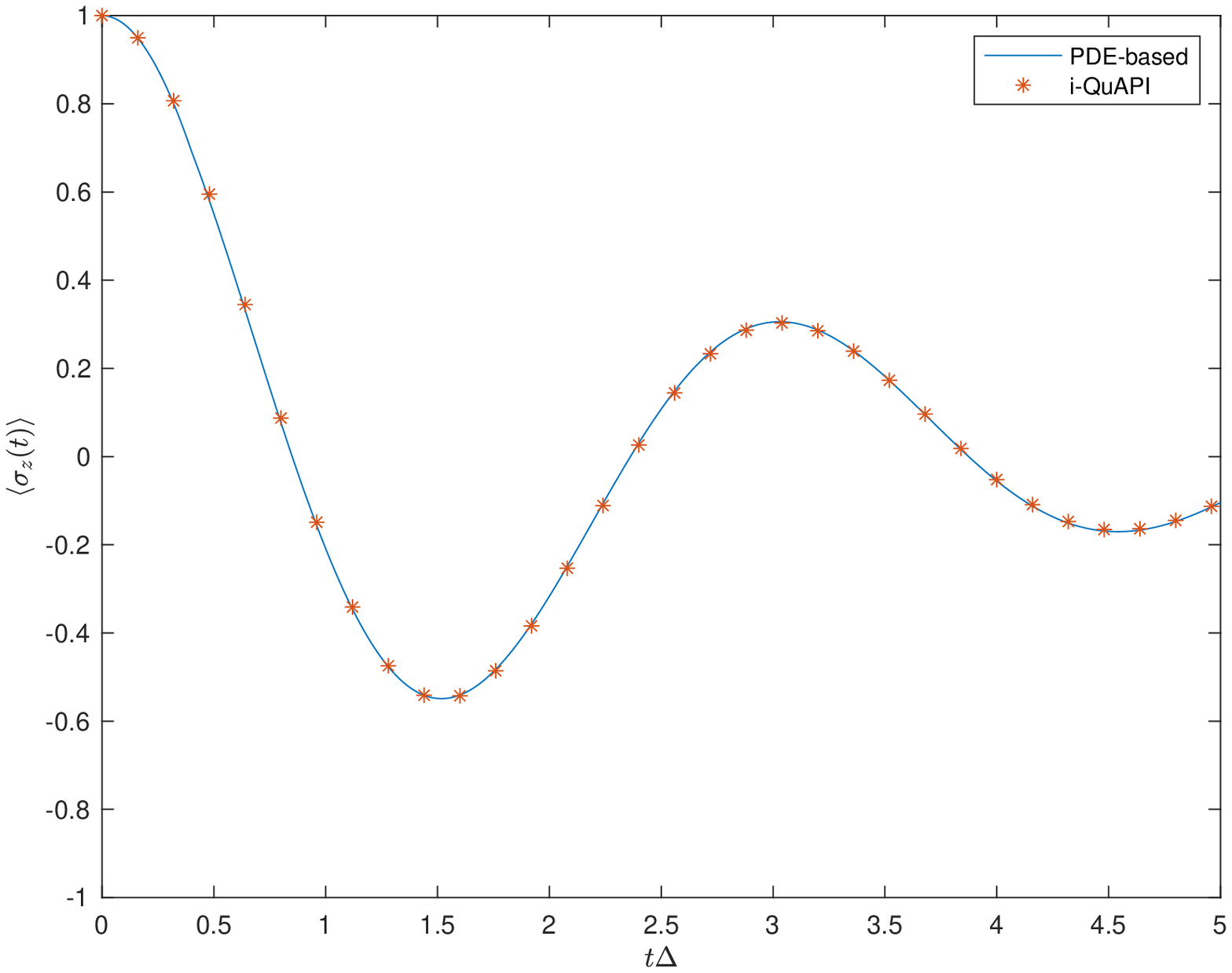}} \quad
    \subfloat[$\beta=5/\Delta$]{\includegraphics[width= 0.3\textwidth]{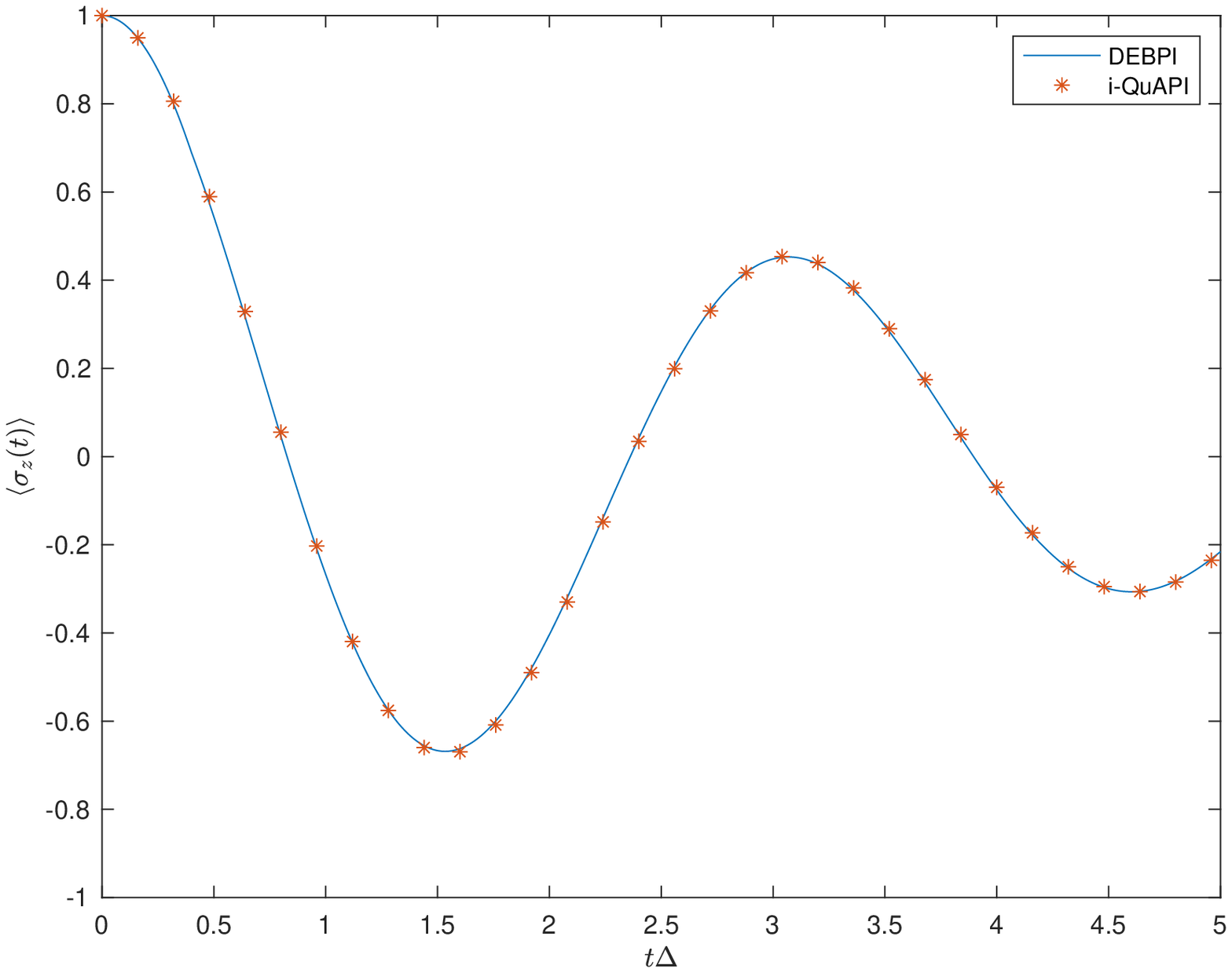}} \quad
    \caption{Both methods use truncation time $T=4$. For i-QuAPI method, $\dk$ is chosen to be 10 and for c-QuAPI, $D_{\max}=3$ and $N=15$.}
    \label{fig_result_data151409}
\end{figure}

For the last set of tests, we choose the parameters
\begin{displaymath}
\xi=0.2, \quad \Delta = 0.1, \quad \omega_c = 2.5\Delta, \quad \epsilon=0,
\end{displaymath}
and we let the inverse temperature of the bath $\beta$ vary from  $0.2/\Delta$ to $5/\Delta$.
The results and the numerical parameters are given in \Cref{fig_result_data151409}. As reflected in the numerical results, in the case of higher temperature, stronger quantum dissipation leads to faster reduction to the state with equal probabilities on both spin states.

Compared with the examples in \Cref{sec:bias}, we use the same memory length $T = 4$, while the spin flipping frequency $\Delta$ is reduced by a half. Thus it can be expected that the value of $D_{\max}$ can also be reduced by a half. Therefore, choosing $D_{\max} = 3$ is sufficient to capture the behavior of evolution processes here.
In these examples, due to the small value of $D_{\max}$, the memory cost of the DEBPI approach is much lower even if $N$ is chosen to be greater than $\dk$. In detail, the DEBPI approach only requires to store $\displaystyle\sum_{D=0}^3 4\times 2^D \begin{pmatrix} D+15 \\ D \end{pmatrix} = 28420$ values, while the i-QuAPI scheme requires $2^{2\times 10} = 1048576$. 

\section{Conclusion and discussion}
\label{Discussion_Conclusion}
We have formulated a PDE system for open quantum systems with truncated memory length in the bath-influence functional.
The PDE system is supplemented with proper initial and boundary conditions, and truncated with a reasonable closure.
Such a formulation allows us to solve the density matrix using classical PDE-solvers for hyperbolic systems.
By this approach, the memory cost can be saved in two possible ways. First, higher-order schemes can be applied to reduce the number of grid points. 
Second, in some situations, we can choose $D_{\max}$ to be smaller than $N$ to avoid discretization of high-dimensional simplicies. 
In our numerical tests, we considered the second-order numerical schemes (the same as i-QuAPI), and illustrated several cases where $D_{\max}$ can be chosen small. 
The higher-order schemes are being studied in our ongoing work.

While this paper focuses mainly on the spin-boson model, 
 the idea of deriving the governing differential equations
 can be generalized to any quantum system
 with finite states without much difficulty.
 When $\mathcal{S}$ contains more elements, the representation of the path segments will be slightly more complicated. In detail, the path segment $h$ cannot be fully determined by the initial state and the locations of state hops. Instead, it is necessary to specify which state the system hops to at each $\tau_k$. Nevertheless, 
 the idea to derive the PDE system and the expression of density matrix stays the same. Roughly speaking, the equation of $A(t,h)$ holds the form 
\begin{equation}
\label{multistate}
  \frac{\partial}{\partial t}A(t,h) = -W(h) A(t,h) + \sum_{\hat{h}} A(t,\hat{h}) + \frac{\partial}{\partial \tau_1} A(t,h),
\end{equation}
where $\tau_1$ is the location of the first state hop in $h$, and the path segment $\hat{h}$ takes all possible paths satisfying
$\hat{h}(\tau) = h(\tau)$ for all $\tau \in (0,T)$ and
\begin{displaymath}
\hat{h}^+(0) = \hat{h}^+(0) \text{ and } \hat{h}^-(0) \neq \hat{h}^-(0) \qquad \text{or} \qquad
\hat{h}^+(0) \neq \hat{h}^+(0) \text{ and } \hat{h}^-(0) = \hat{h}^-(0).
\end{displaymath}
The equation \eqref{multistate} is essentially the same as \eqref{PDEoriginal} in the spin-boson case.
However, when the cardinality of $\mathcal{S}$ is larger, the memory cost also grows faster with $D_{\max}$ and $N$, requiring more efficient numerical methods for the simulation. This will be considered in our future works.

\section*{Appendix. Derivation of i-QuAPI Method}
A component of the reduced density matrix $\rho_s$
 at time $t=N\dt$,
 denoted by $\mel{s''}{\rho_s(N\dt)}{s'}$,
 can be represented by
\begin{equation*}
    \mel{s''}{\rho_s(N\dt)}{s'}
    = \tr_b \mel{s''}{\e^{-\ii H N\dt} \rho(0) \e^{\ii H N \dt}}{s'}.
\end{equation*}
With the completeness relation in a quantum system
\begin{equation*}
	\int_{\mathcal{S}} \dyad{s} \dif s = \id_s,
	\label{completeness}
\end{equation*}
 with $\mathcal{S}$ being the set of
 an orthonormal basis,
 the reduced density matrix can be written as a path integral form
\begin{equation}
    \begin{split}
    \mel{s''}{\rho_s(N\dt)}{s'}
	=& \tr_b \int_{\mathcal{S}} \dif s_0^+ \cdots \int_{\mathcal{S}} \dif s_{N-1}^+
	 \int_{\mathcal{S}} \dif s_0^- \cdots \int_{\mathcal{S}} \dif s_{N-1}^+
	\mel{s''}{\e^{-\ii H \dt}}{s_{N-1}^+} \notag \\
	&\mel{s_{N-1}^+}{\e^{-\ii H \dt}}{s_{N-2}^+} \cdots
	\mel{s_1^+}{\e^{-\ii H \dt}}{s_0^+}
	\mel{s_0^+}{\rho(0)}{s_0^-} \notag \\
	&\mel{s_0^-}{\e^{\ii H \dt}}{s_1^-} \cdots 
	\mel{s_{N-2}^+}{\e^{\ii H \dt}}{s_{N-1}^-}
	\mel{s_{N-1}^-}{\e^{\ii H \dt}}{s'}.
    \end{split}
    \label{pathIntegral}
\end{equation}
\eqref{pathIntegral} is called path integral
 as the states $s_0^+,\cdots,s_N^+$ and $s_0^-,\cdots,s_N^-$
 can be regarded as a path,
 an integral over all possible paths are taken here.
It is clear that the path has two branches 
 (positive branch and negative branch).
 
For the spin-boson model,
 $\mathcal{S}= \{-1,+1\}$,
 therefore, \Cref{pathIntegral} becomes
\begin{align*}
	\mel{s''}{\rho_s(N\dt)}{s'}
	=& \tr_b 
	\sum_{s_0^+=\pm 1} \cdots
	\sum_{s_{N-1}^+ = \pm 1}
	\sum_{s_0^-=\pm 1} \cdots
	\sum_{s_{N-1}^- = \pm 1}
	\mel{s''}{\e^{-\ii H \dt}}{s_{N-1}^+} \notag \\
	&\mel{s_{N-1}^+}{\e^{-\ii H \dt}}{s_{N-2}^+} \cdots
	\mel{s_1^+}{\e^{-\ii H \dt}}{s_0^+}
	\mel{s_0^+}{\rho(0)}{s_0^-} \notag \\
	&\mel{s_0^-}{\e^{\ii H \dt}}{s_1^-} \cdots 
	\mel{s_{N-2}^+}{\e^{\ii H \dt}}{s_{N-1}^-}
	\mel{s_{N-1}^-}{\e^{\ii H \dt}}{s'}.
	\label{pathIntegralSpinBoson}
\end{align*}

As for the special Hamiltonian given by (\ref{Hamiltonian}),
Trotter splitting can be used
 by introducing the following reference Hamiltonian
\begin{equation}
	H_0 = \left(\epsilon \hat{\sigma}_z
    + \Delta \hat{\sigma}_x
    - \sum_j \frac{(c_j \hat{\sigma}_z)^2}{2 \omega_j^2} \right)\otimes \id_b.
	\label{referenceHamiltonian}
\end{equation}

Therefore, the remaining part is
\begin{equation*}
	H - H_0 = \sum_j \left(
	\id_s\otimes 
	\frac{\hat{p}_j^2}{2}
	+ \frac{1}{2} \omega_j^2 
	\left(\id_s\otimes\hat{q}_j
	- \frac{c_j}{\omega_j^2}
	\hat{\sigma}_z \otimes \id_b
	\right)^2 \right).
	\label{H-H_0}
\end{equation*}

When the splitting is applied,
 the reduced density matrix
 can be written as
\begin{align*}
	\mel{s''}{\rho_s(N\dt)}{s'}
	=&\sum_{s_0^+=\pm 1} \cdots
	\sum_{s_{N-1}^+ = \pm 1}
	\sum_{s_0^-=\pm 1} \cdots
	\sum_{s_{N-1}^- = \pm 1}
	\mel{s''}{\e^{-\ii H \dt}}{s_{N-1}^+} \notag \\
	&\mel{s_{N-1}^+}{\e^{-\ii H \dt}}{s_{N-2}^+} \cdots
	\mel{s_1^+}{\e^{-\ii H \dt}}{s_0^+}
	\mel{s_0^+}{\rho_s(0)}{s_0^-} \notag \\
	&\mel{s_0^-}{\e^{\ii H \dt}}{s_1^-} \cdots 
	\mel{s_{N-2}^+}{\e^{\ii H \dt}}{s_{N-1}^-}
	\mel{s_{N-1}^-}{\e^{\ii H \dt}}{s'} \\
	&F(s_0^+, s_1^+, \cdots, s_N^+, s_0^-, s_1^-, \cdots, s_N^-)
	\label{rho1}
\end{align*}
with $s_N^+ = s'', s_N^- = s'$ and 
\begin{equation*}
\begin{split}
	&F(s_0^+, s_1^+, \cdots, s_N^+, s_0^-, s_1^-, \cdots, s_N^-) \\
	=& \tr_b \left[ \e^{-\ii \frac{\dt}{2} (H - H_0(s''))}
	\e^{-\ii \dt (H - H_0(s_{N-1}^+))} \cdots 
	\e^{-\ii \dt (H - H_0(s_1^+))}   \right.  \notag \\
	&\e^{-\ii \frac{\dt}{2} (H - H_0(s_0^+))}
	\rho_b(0)
	\e^{\ii \frac{\dt}{2} (H - H_0(s_0^-))}
	\e^{\ii \dt (H - H_0(s_1^-))} \cdots   \notag \\
	&\left. \e^{\ii \dt (H - H_0(s_{N-1}^-))}
	\e^{\ii \frac{\dt}{2} (H - H_0(s'))}
	\right].
	\end{split}
\end{equation*}
The function $F$ is called influence functional,
 calculating the influence of the environment
 on the system of interest.
With (\ref{referenceHamiltonian})
 as reference Hamiltonian,
 the influence functional can be computed 
 analytically \cite{feynman1963theory}\cite{makri1998quantum}
 by
\begin{equation}
	F(s_0^+, \cdots, s_N^+, s_0^-, \cdots, s_N^-)
	= \exp \left( -\sum_{j_1=0}^{N} \sum_{j_2=0}^{j_1}
	 (s_{j_1}^+ - s_{j_1}^-)
	  (\eta_{j_1,j_2} s_{j_2}^+ - \eta_{j_1,j_2}^* s_{j_2}^-)\right).
	\label{influenceFunctional}
\end{equation}
In the influence functional (\ref{influenceFunctional}),
 $\eta$ is a complex function and $\eta^*$ is its complex conjugate.
For different $j_1,j_2$,
 $\eta$ has different forms \cite{makri1995tensor1}.
For example, when $0 < j_2 < j_1 < N$, we have
\begin{equation*}
	\eta_{j_1,j_2} = 
	\frac{2}{\pi}
	\int_{-\infty}^{\infty} \dif \omega 
	\frac{J(\omega)}{\omega^2} 
	\frac{\exp(\omega\beta/2)}{\sinh(\omega\beta/2)}
	\sin^2 (\omega \dt / 2)
	\exp \left(-\ii (j_1-j_2) \omega \dt \right)
\end{equation*}
where
\begin{equation*}
	J(\omega) = \frac{\pi}{2}\sum_{j}\frac{c_j^2}{m_j\omega_j}\delta(\omega-\omega_j)
\end{equation*}
and $\delta$ is the Dirac delta function.

The absolute value of $\eta$ 
 decreases to zero as $j_1-j_2$ increases.
Thus, when the difference of $j_1$ and $j_2$ is large,
 the contribution of $\eta_{j_1,j_2}$ is small.
From this point of view, the influence functional can be approximated by
\begin{equation*}
	F(s_0^+, \cdots, s_N^+, s_0^-, \cdots, s_N^-)
	\approx \exp \left( -\sum_{j_1=0}^{N} \sum_{j_2=\max\{0,k-\dk\}}^{j_1}
	 (s_{j_1}^+ - s_{j_1}^-)
	  (\eta_{j_1-j_2} s_{j_2}^+ - \eta_{j_1-j_2}^* s_{j_2}^-)\right).
	\label{TruncatedInfluenceFunctional}
\end{equation*}
For simplicity, the notation $S_j = ( s_j^+, s_j^- )$ is introduced and 
the notation $I_{(S_j,S_{j'})}$ is defined by
\begin{equation*}
	I(S_j, S_{j'}) = \begin{cases}
		\e^{- (s_j^+ - s_j^-)(\eta_{j,j'}s_{j'}^+ - \eta_{j,j'}^* s_{j'}^-)},\quad j-j'\not= 1 \\
		\mel{s_j^+}{\e^{-\ii H_0 \dt}}{s_{j-1}^+}
		\mel{s_{j-1}^-}{\e^{\ii H_0\dt}}{s_j^-}
		\e^{- (s_j^+ - s_j^-)(\eta_{j,j'}s_{j'}^+ - \eta_{j,j'}^* s_{j'}^-)},\quad j-j' = 1 
	\end{cases}.
\end{equation*}

With these notations, 
\begin{equation}
    \begin{split}
        &\tilde{\rho}(S_N;N\dt)  \\
        =& \sum_{S_{N-1}}
        \cdots
        \sum_{S_0}
        \prod_{j_1=0}^{N}
        \prod_{j_2=0}^{j_1}
        I(S_{j_1},S_{j_2})
        \mel{s_0^+}{\rho_s(0)}{s_0^-} \\
        =& \sum_{S_{N-1}}
        \cdots
        \sum_{S_0}
        \left(\prod_{j=\dk}^{N}
        \prod_{m=0}^{\dk}
        I(S_{j},S_{j-m})\right)
        \underbrace{\left(
        \prod_{k_1=0}^{\dk-1}
        \prod_{k_2=0}^{k_1}
        I(S_{j_1},S_{j_2})
        \mel{s_0^+}{\rho_s(0)}{s_0^-} \right)}_{:=A_0(S_{\dk-1},\cdots,S_0)} \\
        =& \sum_{S_{N-1}} \cdots \sum_{S_1}
        \left(\sum_{j=\dk+1}^N \sum_{m=0}^k I(S_j,S_{j-m})\right)
        \underbrace{\sum_{S_0}
        \underbrace{\left(\prod_{m=0}^{\dk} I(S_{\dk},S_{\dk-m})\right)}_{:=\Lambda(S_{\dk},\cdots,S_0)}
        A_0(S_{\dk-1},\cdots,S_0)}_{:=A_1(S_{\dk},\cdots,S_1)} \\
        =& \sum_{S_{N-1}} \cdots \sum_{S_2}
        \left(\sum_{j=\dk+2}^N \sum_{m=0}^k I(S_j,S_{j-m})\right)
        \underbrace{\sum_{S_1}
        \underbrace{\prod_{m=0}^{\dk}I(S_{\dk+1},S_{\dk+1-m})}
        _{:=\Lambda(S_{\dk+1},\cdots,S_1)}
        A_1(S_{\dk},\cdots,S_1)
        }_{:=A_2(S_{\dk+1},\cdots,S_2)}\\
        =& \cdots
    \end{split}
    \label{step-by-step}
\end{equation}

An iterative method (i-QuAPI) is designed \cite{makri1995numerical}\cite{makri1998quantum} 
according to \eqref{step-by-step}
as follows.
\begin{align*}
	&A_0(S_{\dk-1}, \cdots, S_0) := \prod_{k_1=0}^{\dk-1} \prod_{k_2=0}^k
		I(S_{k_1}, S_{k_2}) \mel{s_0^+}{\rho_s(0)}{s_0^-};\\
	&\Lambda(S_k,\cdots, S_{k-\dk}) := \prod_{m=0}^{\dk} I(S_k, S_{k-m});\\
	&A_{k+\dk+1} (S_{k},\cdots, S_{k-\dk+1})
	:= \sum_{S_{k-\dk}}
	\Lambda(S_k,\cdots, S_{k-\dk})
	A_{k-\dk}(S_{k-1},\cdots,S_{k-\dk}).
\end{align*}
The density matrix then can be obtained component-wise by
\begin{equation*}
	\mel{s_N^+}{{\rho}_s(N\dt)}{s_N^-}
	= \sum_{S_{N-\dk+1},\cdots,S_{N-1}} A_{N-\dk} (S_{N},\cdots,S_{N-\dk+1}).
\end{equation*}
With a density matrix of the system,
 the expected value of any observable $O$ of the system
 can be computed by
\begin{equation*}
	\expval{O(t)} = \tr(\rho_s(t)O).
	\label{observableOriginal}
\end{equation*}
In spin-boson model,
 the density matrix $\rho_s(t)$ is a two-by-two matrix
 and the observable $O$ is also a two-by-two matrix.

\bibliographystyle{abbrv}
\bibliography{references}

\end{document}